\documentstyle[12pt,epsf]{article}
 \makeatletter
 \def\leqq{\mathrel{\mathpalette\gl@align<}}
 \def\geqq{\mathrel{\mathpalette\gl@align>}}
 \def\gl@align#1#2{\lower.6ex\vbox{\baselineskip\z@skip\lineskip\z@
     \ialign{$\m@th#1\hfil##\hfil$\crcr#2\crcr=\crcr}}}
 \makeatother

 \makeatletter
 \def\sileqq{\mathrel{\mathpalette\gs@align<}}
 \def\sigeqq{\mathrel{\mathpalette\gs@align>}}
 \def\gs@align#1#2{\lower.6ex\vbox{\baselineskip\z@skip\lineskip\z@
     \ialign{$\m@th#1\hfil##\hfil$\crcr#2\crcr\sim\crcr}}}
 \makeatother
\begin{document}
\hbadness=10000
\hbadness=10000
\begin{titlepage}
\nopagebreak
%\def\thefootnote{\fnsymbol{footnote}}
%%%%%%%%%%%%%%%%%%%%%%%%%% Preprint No. %%%%%%%%%%%
\begin{flushright}
%~\\
%~\\
%~\\
%\end{flushright}
{\normalsize
HIP-1999-14/TH\\
KANAZAWA-99-05\\
March, 1999}\\
\end{flushright}
%%%%%%%%%%%%%%%%%%%%%%%%%%%%%%%%%%%%%%%%%%%%%%%%%%%%
\vspace{0.5cm}
\begin{center}

{\large \bf  Electroweak symmetry breaking and s-spectrum in M-theory}

\vspace{0.8cm}
 
{\bf Tatsuo Kobayashi $^{a,b}$
%\footnote[1]{e-mail: kobayash@rock.helsinki.fi}
}, 
{\bf  Jisuke Kubo $^{c}$
%\footnote[2]{e-mail: jik@hep.s.kanazawa-u.ac.jp}
} 
and 
{\bf Hitoshi Shimabukuro $^c$
%\footnote[3]{e-mail: simabuku@hep.s.kanazawa-u.ac.jp}
}

\vspace{0.5cm}
$^a$ Department of Physics, High Energy Physics Division\\
    University of Helsinki, \\
$^b$ Helsinki Institute of Physics\\
    P.O. Box 9, FIN-00014 Helsinki, Finland \\
and\\
$^c$ Institute for Theoretical Physics, Kanazawa University\\
Kanazawa, 920-1192, Japan\\

\end{center}
\vspace{0.5cm}

\nopagebreak
%%%%%%%%%%%%%%%%%%%%%%%%%%%%%

\begin{abstract}
We study soft SUSY breaking parameters in M-theory with and 
without 5-brane moduli fields.
We investigate successful electroweak symmetry breaking and 
the positivity of stau mass squared.
We study mass spectra in allowed regions.
\end{abstract}
%PACS number(s):04.65.+e, 11.25.Mj, 12.60.Jv\\
%Keyword(s):SUSY GUT, coupling reduction,
%soft SUSY breaking parameters, radiative electroweak breaking
\vfill
\end{titlepage}
\pagestyle{plain}
\newpage
\def\thefootnote{\fnsymbol{footnote}}

\section{Introduction}

Supersymmetry (SUSY) breaking is very important to 
connect superstring theory to low-energy physics.
Induced soft SUSY breaking terms are significant to currently running 
and future experiment, and they also have cosmological 
implications.
Actually, soft SUSY breaking terms have been derived, e.g. 
within the framework of weakly coupled heterotic string theory 
under the assumption that only $F$-terms of the dilaton $S$ 
and moduli fields $T^i$ 
contribute to SUSY breaking \cite{CCM,ST-soft,BIM,multiT}.
Also various phenomenological aspects of these string-inspired soft 
terms have been studied.

Recently strongly coupled superstring theories as well as 
strongly coupled SUSY gauge theories have been studied.
Strongly coupled $E_8 \times E_8$ heterotic string theory 
has been considered as M-theory compactified on $S^1/Z_2$ \cite{M-theory}.
In particular, the 4-dimensional effective theory has been obtained 
and supersymmetry (SUSY) breaking has been 
studied \cite{banks}-\cite{ELP}.
M-inspired soft SUSY breaking parameters have a certain  
difference from those derived from 
weakly coupled string models.
Soft parameter formulae between the weakly and strongly coupled 
regions are connected by the parameter $\alpha(T+\bar T)$.
Actually, several phenomenological aspects of soft 
SUSY breaking parameters have been investigated, e.g. radiative 
electroweak symmetry breaking and s-spectrum \cite{BKL}, 
although the large $\tan \beta$ region has not been studied extensively.

The standard embedding leads to the gauge group $E_6 \times E_8'$, 
where the top, bottom and tau Yukawa couplings are unified, 
that is, we have a large value of $\tan \beta$.
In this sense, the large $\tan \beta$ scenario is one natural target 
to study.
When $E_6$ breaks into the standard model (SM) gauge group 
$G_{SM}=SU(3)_C\times SU(2)_L \times U(1)_Y$, another type of soft scalar 
terms appear, i.e., $D$-term contributions.

On the other hand, non-standard embedding including 
Wilson line breaking is also possible \cite{nonst} and that 
can lead to smaller gauge groups, e.g. $SU(3)\times SU(2) \times U(1)$ 
or $SU(5)$.
Thus, in non-standard embedding a small value of $\tan \beta$ can 
also be realized.
Furthermore, non-standard embedding, in general, has 
5-branes, that is, in 4-dimensional effective field theory we have 
a new type of 
moduli fields $Z^n$ whose vacuum expectation values (VEVs) correspond to 
positions of 5-branes $z^n$ within $[0,1]$ along the orbifold $S^1/Z_2$.
It is possible that $F$-terms of the 5-brane moduli fields $Z^n$ also 
contribute to SUSY breaking.
Recently 4-dimensional effective theory with $Z^n$ has been 
studied \cite{5brane} 
and SUSY breaking terms have also investigated \cite{5b-soft}.

The purpose of this paper is to take into account several effects which 
are mentioned above and to investigate successful electroweak breaking 
conditions and the lightest superparticle (LSP) systematically.
In the whole paper, we assume that the massless particle content 
below the grand unified theory (GUT) scale is exactly the same as 
the minimal supersymmetric standard model (MSSM).
In the large $\tan \beta$ scenario, the stau (mass)$^2$ becomes small and 
negative because of large and negative radiative corrections from 
the large Yukawa coupling.
Thus the stau mass leads to a significant constraint for 
large $\tan \beta$.
We cover the whole realistic region of $\tan \beta$ including 
large $\tan \beta$.
We investigate $\alpha(T+\bar T)$-dependence of  
electroweak symmetry breaking and s-spectrum.
Furthermore, we study the effect due to $F$-terms 
of the moduli fields $Z^n$.

This paper is organized as follows.
In section 2, after we describe the structure of the soft parameters 
in M-theory without 5-brane, 
we investigate the electroweak symmetry breaking conditions and 
the positivity of stau mass squared under the assumption of 
top-bottom-tau Yukawa coupling unification, that is, the large 
$\tan \beta$ scenario. 
$D$-term contributions, which appear in $E_6$ breaking into $G_{SM}$, 
are also taken into account.
We also consider the case with small $\tan \beta$.
In section 3, we consider soft parameters in M-theory that  
the 5-brane moduli field also contribute to SUSY breaking.
In this case, electroweak symmetry breaking and mass spectra are studied 
and allowed regions are compared with the case without 5-brane effects.
Section 4 is devoted to conclusions.

\section{SUSY breaking in M-theory without 5-branes}
In this section we study SUSY breaking in M-theory without 
5-branes.
For example, standard embedding corresponds to this class.
We discuss the soft breaking parameters derived within this framework 
and study their phenomenological implications.

\subsection{Soft parameters}

The 4-dimensional effective supergravity action of
M-theory can be analyzed by expanding it in powers of the two
dimensionless variables:
$\varepsilon_1=\kappa^\frac{2}{3}\pi\rho/V^\frac{2}{3}$ and
$\varepsilon_2=\kappa^\frac{2}{3}/\pi\rho V^\frac{1}{3}$, likewise
the weakly coupled heterotic string theory expanded by the string
coupling $\varepsilon_s=e^{2\phi}/(2\pi)^5$ and the worldsheet sigma
model coupling $\varepsilon_\sigma=4\pi\alpha'/V^{1/3}$, 
where $\kappa^2$, $\rho$ and $V$ denote the 11-dimensional gravitational 
coupling, the 11-dimensional length and 
the Calabi-Yau volume respectively.

Here we consider only the overall moduli field $T$.
Thus, the 4-dimensional effective supergravity includes 
the dilaton $S$ and the moduli field $T$ in addition to 
chiral multiplets $\Phi^p$, gauge multiplets and the graviton 
multiplet.
Scalar components of $S$ and $T$ can be identified as
\begin{eqnarray}
Re(S)&=&\frac{1}{2\pi}(4\pi \kappa^2)^{-2/3}V \nonumber ,\\
Re(T)&=&\frac{6^{1/3}}{8\pi}(4\pi \kappa^2)^{-1/3}\pi\rho V^{1/3}.
\end{eqnarray}

The K\"{a}hlar potential is computed by M-theory
expansion for matter fields up to order of $\kappa^{4/3}$ \cite{LOW},
\begin{eqnarray}
K&=&-\log(S+\bar{S})-3\log(T+\bar{T})+\left(\frac{3}{T+\bar{T}}
   +\frac{\alpha}{S+\bar{S}}\right)|\Phi|^2 \label{K}.
\end{eqnarray}
Similarly the gauge kinetic functions of the observable and hidden sectors 
are calculated \cite{banks,choi1,nilles1,nilles2},
\begin{eqnarray}
f_{E_6}&=&S+\alpha T, \hspace{6pt} f_{E_8}=S-\alpha T  \label{f6},
\end{eqnarray}
where $\alpha={1}/{(8\pi^2)}\int \omega \wedge \left[tr(F
\wedge F)-\frac{1}{2}tr(R \wedge R)\right]$.
In addition, the tree-level superpotential is obtained, 
\begin{eqnarray}
W&=&Y_{pqr}\Phi^p\Phi^q\Phi^r \label{W}.
\end{eqnarray}
Here $\alpha$ is of the order 
one at the ``physical'' point where gauge coupling unification can be
realized \cite{banks}.
Note that the superpotential $W$ and gauge
kinetic functions $f_{E_6}$ and $f_{E_8}$ are exact up to nonperturbative
corrections, while there are small additional perturbative
corrections to the K\"{a}hlar potential which are of order
$1/4\pi^2Re(S)$ or $1/[4\pi^2Re(T)]^3$ $\sim O(\alpha_{GUT}/\pi)$ in
the strong coupling limit $\varepsilon_s \gg 1$.

Now we know the forms of the K\"ahler potential $K$, the gauge kinetic 
function $f_{E_6}$ and the tree-level superpotential $W$.
Thus, we can derive the formula of soft SUSY breaking parameters 
following ref.\cite{SW,ST-soft}.
We assume that a nonperturvative superpotential of $S$ and $T$ is induced and 
$F$-terms of $S$ and $T$ contribute to SUSY breaking.

The gravitino mass $m_{3/2}$ is obtained 
\begin{equation}
3m_{3/2}^2 = \frac{|F^S|^2}{3(S+\bar{S})^2}
+\frac{|F^T|^2}{(T+\bar{T})^2}-V_0,
\end{equation}
where $V_0$ is the vacuum energy.
Hence, following ref.\cite{BIM}, we parameterize $F$-terms,
\begin{eqnarray}
F^S&=&\sqrt{3}m_{3/2}C(S+\bar{S})\sin\theta e^{-i\gamma_S} \label{Fs},\\
F^T&=&m_{3/2}C(T+\bar{T})\cos\theta e^{-i\gamma_T},\label{Ft}
\end{eqnarray}
where $C^2=1+V_0/3m_{3/2}$. 
Using these parameters, we can write the soft parameters, i.e., 
the gaugino mass $M_{1/2}$, the soft scalar mass $m$ and the 
$A$-parameter as follows \cite{CKM}
\begin{eqnarray}
M_{1/2}&=&\frac{\sqrt{3}Cm_{3/2}}{(S+\bar{S})+\alpha(T+\bar{T})}
\left((S+\bar{S})
   \sin\theta e^{-i\gamma_S} \right. \nonumber \\
&+& \left. \frac{\alpha(T+\bar{T})}{\sqrt{3}}\cos\theta 
   e^{-i\gamma_T}\right) \label{Mg},\\
m^2&=&V_0+m^2_{3/2}-\frac{3C^2m^2_{3/2}}{3(S+\bar{S})+\alpha(T+\bar{T})} 
     \nonumber\\
&{}& \times 
 \left\{
    \alpha(T+\bar{T})\left(2-\frac{\alpha(T+\bar{T})}{3(S+\bar{S})
       +\alpha(T+\bar{T})}\right)\sin^2\theta \right.  \nonumber\\
&{}& \left. +(S+\bar{S})\left(2-\frac{3(S+\bar{S})}{3(S+\bar{S})
       +\alpha(T+\bar{T})}\right)\cos^2\theta \right.\nonumber\\
&{}& \left. -\frac{2\sqrt{3}\alpha(T+\bar{T})(S+\bar{S})}{3(S+\bar{S})
       +\alpha(T+\bar{T})}\sin\theta\cos\theta\cos(\gamma_S-\gamma_T)
 \right\} \label{ms},\\
A&=&\sqrt{3}Cm_{3/2}
 \left\{
    \left(-1+\frac{3\alpha(T+\bar{T})}{3(S+\bar{S})
       +\alpha(T+\bar{T})}\right)\sin\theta e^{-i\gamma_S} \right.\nonumber\\
&{}& \left. +\sqrt{3}\left(-1+\frac{3(S+\bar{S})}{3(S+\bar{S})
       +\alpha(T+\bar{T})}\right)\cos\theta e^{-i\gamma_T}
 \right\}. \label{A}
\end{eqnarray}
In following analyses 
we concentrate to the case with the vanishing vacuum energy, $V_0=0$ 
($C=1$) and no CP phase $\gamma_S=\gamma_T=0$.

Since the gauge kinetic function $f_{E_6}$ provides the gauge 
coupling constant, that is, $Ref_{E_6}=1/g^2_{E_6}$, 
the phenomenological requirement $g^{-2}_{GUT} \simeq 2$ leads to
\begin{eqnarray}
(S+\bar{S})+\alpha(T+\bar{T}) \simeq 4 \label{cons}.
\end{eqnarray}
Moreover, the constraint for $Ref_{E_8} > 0$ leads to $(S+\bar{S})
>\alpha(T+\bar{T})$.
Therefore, we have 
\begin{eqnarray}
 0 < \alpha(T+\bar{T}) \sileqq 2 \label{lam}.
\end{eqnarray}
Note that there are three free parameters $m_{3/2}$, $\theta$ and
$\alpha(T+\bar{T})$, and in particular, 
$\alpha(T+\bar{T})$ is the characteristic parameter of M-theory.
The limit $\alpha(T+\bar{T}) \to 0$ corresponds to weakly coupled theory.

\subsection{Electroweak symmetry breaking and mass spectrum}

We now analyze phenomenology of M-theory by using
eqs.(\ref{Mg}),(\ref{ms}),(\ref{A}) as boundary conditions of
renormalization group equations (RGEs) 
at the GUT scale.
We investigate successful radiative electroweak symmetry breaking and 
positive masses squared for any sfermion.
That gives constraints for the parameters $m_{3/2}$, $\alpha(T+\bar{T})$ 
and $\theta$.

If we assume a certain type of $\mu$-term generation mechanism, 
we could fix magnitudes of the supersymmetric Higgs mixing mass $\mu$ and 
the corresponding SUSY breaking parameter $B$.
However, we do not take such a procedure here.
Because we would like to study generic case.
We determine these magnitudes by using the following 
minimization conditions of the MSSM Higgs potential,
\begin{eqnarray}
 m_1^2+m_2^2&=&-\frac{2 \mu B}{\sin\beta}, \nonumber \\
 m_1^2-m_2^2&=&-(M_Z^2+m_1^2+m_2^2)\cos2\beta ,\label{min}
\end{eqnarray}
where $m_{1,2}^2=m_{H_1,H_2}^2+\mu^2$, and $m_{H_1,H_2}^2$ are soft scalar
masses of Higgs fields.
The Higgs fields for the down sector and the up sector are denoted by 
$H_1$ and $H_2$. 
Using these equations, $\mu$ and $B$ are written in terms of 
the soft scalar masses and $\tan \beta$.

Realization of successful electroweak symmetry 
breaking leads to the condition,
\begin{eqnarray}
m_1^2 m_2^2 < |\mu B|^2, \label{EW}
\end{eqnarray}
and the bounded-from-below condition along the $D$-flat
directions in the Higgs potential, that is, 
\begin{eqnarray}
m_1^2+m_2^2 >  2|\mu B|,
\end{eqnarray}
should be satisfied in order for stability of the potential 
along the $D$-flat direction. 
In the large $\tan \beta$ scenario, these conditions lead to 
\begin{eqnarray}
m_{H_1}^2-m_{H_2}^2 \sigeqq M_Z^2/2 .\label{EW2} 
\end{eqnarray}

Also, we require that a physical mass squared should be positive 
for any sfermion.
Since the sleptons have no large positive-definite radiative
corrections from loops involving the gluino, 
the sleptons become lighter than squarks at low energy.
In particular, in the case of large tan$\beta$, the stau mass
squared has a sizable and negative radiative correction due to 
the large Yukawa coupling.
lighter stau mass becomes significant 
for large $\tan \beta$ \footnote{ For example, in refs.\cite{kkmz,kks}
the constraint  $m_{\tilde{\tau}_1}^2>0$ has been considered 
in details for (finite) $SU(5)$ GUTs and $SO(10)$ GUTs.}.
Furthermore, what is the LSP is an important issue.
One candidate is the lightest neutralino and another candidate is 
the lighter stau for large $\tan \beta$ 
from the above reason \footnote{In ref.\cite{stau} cosmological 
implications of the stau LSP have been discussed.
The stau LSP could not be a candidate for cold dark matter.
In the case with the stau LSP we need another candidate 
for cold dark matter.}.
The stau mass matrix is
\begin{eqnarray}
{\bf m_{\tilde{\tau}}^2} = 
\left(
\begin{array}{cc}
m_{\tilde{\tau}_L}^2+(-\frac{1}{2}+\sin^2\theta_W)M_Z^2 \cos2\beta
& vY_\tau(A_\tau \cos\beta-\mu \sin\beta) \\
vY_\tau(A_\tau \cos\beta-\mu \sin\beta) 
& m_{\tilde{\tau}_R}^2-M_Z^2 \sin^2\theta_W \cos2\beta
\end{array}
\right) \label{taumtrx},
\end{eqnarray}
where $v^2 \equiv <H_1>^2+<H_2>^2$, and we have neglected the tau
(mass)$^2$ $M_\tau^2$ term here.
The lighter stau mass squared is obtained as 
\begin{eqnarray}
m_{\tilde{\tau}_1}^2&=&\frac{1}{2}
\Biggl(
m_{\tilde{\tau}_L}^2+m_{\tilde{\tau}_R}^2
-\frac{1}{2}M_Z^2 \cos2\beta \nonumber \\
&-& [(m_{\tilde{\tau}_L}^2-m_{\tilde{\tau}_R}^2 -(\frac{1}{2}-2\sin^2
\theta_W) M_Z^2 \cos2\beta )^2 \nonumber \\
&+& 4vY_\tau(A_\tau \cos\beta-\mu \sin\beta)^2]^{1/2}
\Biggr).  \label{tau1}
\end{eqnarray}

We take the following input parameters,
\begin{eqnarray}
M_\tau&=&1.777GeV,\hspace{12pt} M_Z=91.188GeV \nonumber, \\
\alpha_{EM}^{-1}(M_Z)&=&127.9+\frac{8}{9\pi}\log\frac{M_t}{M_Z} \nonumber, \\
\sin^2\theta_W&=&0.2319-3.03 \times 10^{-5}\Delta_t-8.4 \times 
10^{-8}\Delta_t^2 ,   
\label{input}
\end{eqnarray}
where $\Delta_t=M_t[GeV]-165.0$. Here $M_\tau$ and $M_t$ are physical tau
lepton and top quark masses.
We define the SUSY scale as the arithmetic average of the stop
squared mass eigenvalues \cite{MSUSY},
\begin{eqnarray}
M_{SUSY}^2=\frac{m_{\tilde{t_1}}^2+m_{\tilde{t_2}}^2}{2} \label{SUSY}.
\end{eqnarray}

\subsubsection{Standard embedding}

First of all, we consider the case of standard embedding, where the 
gauge group $E_6 \times E_8'$ is obtained and the top, bottom and tau Yukawa 
couplings are unified at the GUT scale. From the experimental value of 
$M_\tau$, we obtain tan$\beta=53$ and $M_t=175$GeV. 
We assume at the GUT scale $E_6$ breaks into $G_{SM}$ and 
exactly the MSSM matter content is obtained.
On top of that, we assume this symmetry breaking induces no extra 
contributions to soft parameters.
Then we use eqs.(\ref{Mg}),(\ref{ms}),(\ref{A}) as the initial conditions 
of RGEs at the GUT scale.
The latter assumption will be relaxed later.

Before studying the constraints, it is useful to show 
how the soft parameters  
depend on the parameters, $2\alpha Re(T)$ and $\theta$.
The $A$-parameter is always negative.
Fig. 1 shows the ratios $-A/M_{1/2}$ and $m/M_{1/2}$ against $2\alpha Re(T)$ 
for values of $\theta$.
The three solid curves show $-A/M_{1/2}$.
The upper, intermediate and lower curves correspond to $-A/M_{1/2}$ for 
$\theta =0, \pi/5$ and $\pi/2$, respectively.
Similarly, the two dotted curves show $m/M_{1/2}$.
The upper and lower correspond to $m/M_{1/2}$ for 
$\theta =\pi/5$ and $\pi/2$, respectively.
The weak coupling limit leads to 
\begin{equation}
{A \over M_{1/2}} =-1, \quad {m \over M_{1/2}} = {1 \over \sqrt 3},
\label{sumrule}
\end{equation}
for any value of $\theta$ \footnote{These relations have an implication 
for finiteness and RG invariance \cite{JJ,kkmz}.}.
We have $-A/M_{1/2} \approx -1$ for any value of $2\alpha Re(T)$ 
except the dilaton-dominant case $|\sin \theta| \approx 1$ and 
the moduli-dominant case $|\cos \theta| \approx 1$.
In the dilaton-dominant (moduli-dominant) case, $-A/M_{1/2}$ becomes 
smaller (larger) in particular at $2\alpha Re(T) > 1.5$.
The ratio $m/M_{1/2}=1/\sqrt 3$ holds approximately for any value of 
$2\alpha Re(T)$ except in the case with $|\sin \theta | > 0.9$ and 
the purely moduli-dominant case with $|\cot \theta| > O(10)$.
In the moduli dominant case with $|\sin \theta | > 0.9$, 
as $2\alpha Re(T)$ increases, $m/M_{1/2}$ becomes small in particular 
at $2\alpha Re(T)> 1.5$.
Thus, we have similar results in the region where the relations 
(\ref{sumrule}) hold approximately.

Figs. 2 show the regions excluded by the electroweak symmetry breaking 
conditions and the constraint $m^2_{\tilde{\tau}_1} > 0$.
The dotted region denotes the forbidden parameter region by the
electroweak breaking conditions, while the squares correspond to 
$m^2_{\tilde{\tau}_1} < 0$.
In addition the asterisks denote the region with  $m^2_{\tilde{\tau}_1} <
m^2_{\tilde{\chi}^0_1}$, where the LSP is the stau.

The constraint due to successful electroweak symmetry breaking is 
relaxed more as $2\alpha Re(T)$ increases.
In particular, a large $\theta$ within $[0,\pi]$ is favorable for successful 
electroweak symmetry breaking.
On the other hand, in the case with a large $\theta$, we 
have $m^2_{\tilde{\tau}_1} < 0$ for large $2\alpha Re(T)$.
That is because  we have $m^2 < 0$ at the GUT scale by
eq.(\ref{ms}), e.g. in the cases with $\theta=\pi/2$ and 
$3\pi/4$ for $\alpha(T+\bar{T}) > 1.8$ and $1.3$, respectively.  
In any allowed region, the LSP is the stau.

\subsubsection{$D$-term contributions}

We have assumed that symmetry breaking has no effect on soft parameters.
However, if a gauge group symmetry breaks reducing its rank, 
additional contributions to soft scalar mass terms, in general, appear, 
i.e. $D$-term contribution \cite{Dterm1,Dterm2}.
Thus, when the gauge group $E_6$ breaks to $G_{SM}$, 
the additional contributions to soft
scalar mass are able to appear at the breaking scale.
Here we take into such $D$-term contributions effects to soft scalar masses.
In general, these $D$-term contributions are proportional to
quantum numbers of broken diagonal generators.  
In the present case, we have two independent universal factors of $D$-term
contributions, which we denote as $m_{D1}^2$ and  $m_{D2}^2$, 
because the rank of $E_6$ is larger than $G_{SM}$ by two. 
%Instead of diterminating to their
%magnitudes by concrete model buildings, due to analyze general model, 
We regard these as free parameters and then analyze their effects 
on the parameter space. 
The soft scalar masses at the GUT scale including $D$-term contributions are
written as 
\begin{eqnarray}
m_{\tilde{Q}}^2=m_{\tilde{t}}^2=m_{\tilde{\tau}_R}^2&=&
m_{27}^2-m_{D1}^2+m_{D2}^2 , \nonumber\\
m_{\tilde{b}}^2=m_{\tilde{\tau}_L}^2&=& 
m_{27}^2+3m_{D1}^2+m_{D2}^2 , \nonumber\\
m_{H_1}^2&=& m_{27}^2-2m_{D1}^2-2m_{D2}^2 , \nonumber\\
m_{H_2}^2&=& m_{27}^2+2m_{D1}^2-2m_{D2}^2 ,\label{Dterm} 
\end{eqnarray}
where $m_{27}^2$ is the soft scalar mass in eq.(\ref{ms}).

We now analyze $D$-term contribution effects on the parameter space 
fixing $\tan \beta =53$, which has been obtained under the assumption of 
the top-bottom-tau Yukawa coupling unification.
The term $m_{D1}^2$ with a negative and sizable value is helpful to 
create a positive gap $m_{H1}^2-m_{H2}^2$, which is favorable to 
successful electroweak symmetry breaking.
Also, a negative value of $m_{D1}^2$ increases $m_{\tilde{\tau}_R}^2$, 
but it reduces $m_{\tilde{\tau}_L}^2$.
The (mass)$^2$ $m^2_{\tilde{\tau}_R}$ is dominant to $m^2_{\tilde{\tau}_1}$ 
around the universal case with  $m^2_{\tilde{\tau}_R}=m^2_{\tilde{\tau}_L}$.
Therefore negative $m_{D1}^2$ may increase $m^2_{\tilde{\tau}_1}$ and 
seems to be favorable to the constraint of $m^2_{\tilde{\tau}_1}$.
However, as $m_{D1}^2$ decreases, $m^2_{\tilde{\tau}_L}$ is decreased  
and below a critical value of $m_{D1}^2$ 
$m^2_{\tilde{\tau}_L}$ becomes dominant to $m^2_{\tilde{\tau}_1}$. 
A similar effect of the $D$-term contribution has been discussed within 
the framework of $SO(10)$ GUTs \cite{kks}.

On the other hand, a positive value of $m_{D2}^2$ increases both 
$m^2_{\tilde{\tau}_R}$ and $m^2_{\tilde{\tau}_L}$.
Thus, a positive value of $m_{D2}^2$ is favorable to increase 
$m^2_{\tilde{\tau}_1}$.
The effects of $m_{D2}^2$ to $m_{H1}^2$ and $m_{H2}^2$ are same.
Hence, $m_{D2}^2$ has little effect on electroweak symmetry breaking.

These results are shown in Figs.3 and Figs. 4 for $\theta = \pi/2$. 
Figs. 3 show the $m_{D2}^2$ effects. 
A large value of $m_{D2}^2$ reduces the region excluded by the 
constraint $m^2_{\tilde{\tau}_1} >0$, 
while it has little effect on electroweak symmetry breaking.
The region with the neutralino LSP appears and it becomes wider 
as $m_{D2}^2$ increases.
For example, in the case of $m_{D2}^2=1.0$ (TeV)$^2$, 
the region with the neutralino LSP has the 
largest value $m_{3/2}=3.1$ TeV when $\alpha(T+\bar{T})=0.8$.
In this case the lightest neutral Higgs boson mass
$m_{h^0}$ is 151 GeV.
Here we use the two loop Higgs mass $m_{h^0}$\cite{MSUSY,HIGS}.
Similarly, we obtain $m_{h^0}=129$ GeV when
$m_{3/2}=1.0$ TeV and $\alpha(T+\bar{T})=0.8$. 
This value is stable between $m_{3/2}=0.4$ TeV and 1.0 TeV.

Figs. 4 show the $m_{D1}^2$ effects. 
They show that the region excluded by the electroweak symmetry breaking 
conditions becomes more narrow and the region with 
$m^2_{\tilde{\tau}_1} <0$ becomes wider as $m_{D1}^2$ decreases.
In suitable values of $m_{D1}^2$, we have the region with 
the neutralino  LSP.
Figs. 4 show the critical value is around $m_{D1}^2=-2.0$ (TeV)$^2$, 
where the neutralino LSP region is widest.

\subsubsection{Small $\tan \beta$ scenario}

It is also possible that we obtain a smaller gauge group, e.g. 
$G_{SM}$ or $SU(5)$ at the GUT scale by non-standard embedding 
including Wilson line breaking.
In this case, the assumption of the top-bottom-tau Yukawa coupling 
unification can be relaxed and 
a smaller value of $\tan \beta$ can be realized.
We assume that below the GUT scale the matter content is 
exactly same as the MSSM.
Here, we vary tan$\beta$ holding eqs.(\ref{Mg}),(\ref{ms}),(\ref{A}) 
as the initial conditions of RGEs at the GUT scale 
and fixing $M_t=175$ GeV.

Figs. 5 show that the allowed parameter region for tan$\beta$.
Fig. 5a  shows that for a small value of $|\sin \theta|$ 
we have no excluded region at $\tan \beta < 50$.
However, for large values of $|\sin \theta|$ and 
$2 \alpha Re(T)$, both the regions with $m^2_{\tilde{\tau}_1} <0$ and 
the stau LSP remain even when $\tan \beta $ is small as shown in 
Fig. 5b.

\section{M-theory with 5-brane}

\subsection{soft parameters}

In general, non-standard embedding includes 5-branes, 
that is, 4-dimensional effective theory includes 5-brane moduli 
fields $Z^n$.
It is possible that $Z^n$ also contribute to SUSY breaking.
Thus, we consider such effects for generic case.
If the 5-branes exist between the boundaries in M-theory, the
K$\ddot{\mbox{a}}$hler potential and 
the gauge kinetic functions  are modified. 
The relevant part of the moduli
K$\ddot{\mbox{a}}$hler potential $K_{mod}$ and the gauge kinetic functions 
of the observable and hidden sectors $f^{(1)}$ and $f^{(2)}$ 
with $N$ 5-branes take the following form \cite{5brane,5b-soft},
\begin{eqnarray}
K_{mod}&=&-ln(S+\bar{S})-3ln(T+\bar{T})+K_5 \label{Kmod}, \\
f^{(1)}&=&S+\epsilon T\left(\beta^{(0)}
           +\sum_{n=1}^{N}(1-Z^n)^2\beta^{(n)}\right),
\nonumber \\
f^{(2)}&=&S+\epsilon T\left(\beta^{(N+1)}
            +\sum_{n=1}^{N}(Z^n)^2\beta^{(n)}\right) \label{5gfunc},
\end{eqnarray}
where $K_5$ is the K$\ddot{\mbox{a}}$hler potential for the 5-brane
position moduli $Z^n$ and  the expansion parameter 
$\epsilon=\left({\kappa}/{4\pi}\right)^{2/3}{2\pi^2\rho}/{V^{2/3}}$.
In addition, 
$\beta^{(0)}$, $\beta^{(N+1)}$ are the instanton numbers on the
observable sector boundary and the hidden sector boundary respectively,
and $\beta^{(n)}$ $(n=1,\cdots,N)$ is a magnetic charge on the each
5-brane.  
They should satisfy the cohomology constraint,
\begin{eqnarray}
\sum_{n=0}^{N+1}\beta_i^{(n)}=0 \label{coho},
\end{eqnarray}
which means physically that the net charges must be zero 
because the ``flux'' cannot get away anywhere in
compact space.

The part of the matter fields is obtained
\begin{eqnarray}
K_{mat}=\left(\frac{3}{(T+\bar{T})}+\frac{\epsilon\zeta}
            {(S+\bar{S})}\right)|\Phi|^2, 
\end{eqnarray}
where 
\begin{eqnarray}
\zeta=\beta^{(0)}+\sum_{n=1}^{N}(1-z^n)^2\beta^{(n)}\label{zeta}.
\end{eqnarray}

Following the generic formulae of refs.\cite{SW,ST-soft} again, 
we can derive the soft parameters.
We assume $V_0=0$ again.
The soft parameters are obtained, 
\begin{eqnarray}
M_{1/2}&=&\frac{1}
                {(S+\bar{S})+\epsilon\zeta(T+\bar{T})}
               \Bigl[F^S+\epsilon\zeta F^T+\epsilon T F^n \zeta_n\Bigr]  \\
m^2&=&m_{3/2}^2
    -|F^S|^2\Bigl[\frac{1}{(S+\bar{S})^2}
       -\frac{9}{\Bigl(3(S+\bar{S})+\epsilon\zeta(T+\bar{T})\Bigr)^2}\Bigr]
 \nonumber \\
   &-&|F^T|^2\Bigl[\frac{1}{(T+\bar{T})^2}
       -\frac{(\epsilon\zeta)^2}
          {\Bigl(3(S+\bar{S})+\epsilon\zeta(T+\bar{T})\Bigr)^2}\Bigr]  
\nonumber \\
   &-&F^n\bar{F}^{\bar{m}}\Bigl[\frac{\epsilon\zeta_{n\bar{m}}(T+\bar{T})}
         {3(S+\bar{S})+\epsilon\zeta(T+\bar{T})}
        -\frac{\epsilon^2\zeta_n \zeta_{\bar{m}}(T+\bar{T})^2}
            {\Bigl(3(S+\bar{S})+\epsilon\zeta(T+\bar{T})\Bigr)^2}\Bigr]
\nonumber \\
   &+&3F^S\bar{F}^{\bar{T}}\frac{\epsilon\zeta}
              {\Bigl(3(S+\bar{S})+\epsilon\zeta(T+\bar{T})\Bigr)^2}
   +3F^S\bar{F}^{\bar{m}}\frac{\epsilon\zeta_{\bar{m}}(T+\bar{T})}
              {\Bigl(3(S+\bar{S})+\epsilon\zeta(T+\bar{T})\Bigr)^2}
\nonumber \\
   &-&F^T\bar{F}^{\bar{m}}\Bigl[\frac{\epsilon\zeta_{\bar{m}}}
         {3(S+\bar{S})+\epsilon\zeta(T+\bar{T})}
      -\frac{\epsilon^2\zeta\zeta_{\bar{m}}(T+\bar{T})}
        {\Bigl(3(S+\bar{S})+\epsilon\zeta(T+\bar{T})\Bigr)^2}\Bigr]+(c.c)
\nonumber \\
&{}&  \\
A&=&F^S\Bigl[\frac{2}{S+\bar{S}}
       -\frac{9}{3(S+\bar{S})+\epsilon\zeta(T+\bar{T})}\Bigr]
\nonumber \\
    &+&F^T\frac{\epsilon\zeta}{3(S+\bar{S})+\epsilon\zeta(T+\bar{T})}
\nonumber \\
    &+&F^n\Bigl[K_{5,n}-\frac{3\epsilon\zeta_n(T+\bar{T})}
         {3(S+\bar{S})+\epsilon\zeta(T+\bar{T})}\Bigr],    
\end{eqnarray}
where $K_{5,n}=\partial K_5/\partial Z^n$, 
$\zeta_{n}=\partial \zeta/\partial z^n$ and 
$\zeta_{nm}=\partial^2 \zeta/(\partial z^n \partial z^m)$.

We have normalized the soft scalar mass as, 
\begin{eqnarray}
m_{IJ}^2=m^2{\partial^2 K_{mat} \over |\partial \Phi |^2},
\end{eqnarray} 
using the total K\"ahler metric.
That is different from ref.\cite{5b-soft}, where 
the soft scalar mass is normalized as,
\begin{eqnarray}
m_{IJ}^2=m^2{3 \over (T +\bar T)^2}.
\end{eqnarray} 
In addition, here we have kept all terms, although in ref.\cite{5b-soft} 
only the linear terms of $\varepsilon$ have been kept.

Here we ignore CP phases, i.e. $F=\bar{F}$.
For simplicity, we assume that there is only one relevant 
5-brane moduli $Z$ and its K\"ahler potential is a function of only 
$(Z+\bar Z)$, i.e., 
\begin{eqnarray}
K_5=K_5(Z+\bar{Z}). \label{asmp1}
\end{eqnarray}

The gravitino mass is obtained, 
\begin{eqnarray}
m_{3/2}^2=\frac{|F^S|^2}{3(S+\bar{S})^2}+\frac{|F^T|^2}{(T+\bar{T})^2}
            +\frac{1}{3}K_{5,Z\bar{Z}}|F^Z|^2\label{5mgra}.
\end{eqnarray}
Now we can parameterize each $F$-term as follows
\begin{eqnarray}
F^S&=&\sqrt{3}m_{3/2}(S+\bar{S})\sin\theta \sin\phi, \nonumber \\
F^T&=&m_{3/2}(T+\bar{T})\cos\theta \sin\phi, \nonumber \\
F^Z&=&\sqrt{\frac{3}{K_{5,Z\bar Z}}}m_{3/2}\cos\phi.
\end{eqnarray}

Still now we have several unknown parameters and most of them are due to 
detailed information of the 5-brane.
Thus, we fix $z$ and $\beta^{(1)}$, e.g. 
\begin{eqnarray}
z=\frac{1}{2},\hspace{1cm} \beta^{(1)}=\frac{4}{3}\beta^{(0)} \label{asmp2}.
\end{eqnarray}
Note that both the instanton number $\beta^{(0)}$ and the intersection 
number $\beta^{(1)}$ are integers. 
Eq.(\ref{asmp2}) leads to the following simple relations with
eq.(\ref{zeta}) as 
\begin{eqnarray}
\zeta|_{z={1}/{2}}=-\zeta_{,z}|_{z={1}/{2}}
            =\frac{1}{2}\zeta_{,zz}|_{z={1}/{2}}
	    =\beta^{(1)} \label{relazeta}.
\end{eqnarray}

In this case we can rewrite the soft parameters,
\begin{eqnarray}
M_{1/2}&=&\frac{m_{{3}/{2}}}{(S+\bar{S})+\epsilon\zeta(T+\bar{T})}
\left\{
       \sqrt{3}(S+\bar{S})
     \sin\theta \sin\phi
\right.  \nonumber \\
   &+&\epsilon\zeta(T+\bar{T})\cos\theta \sin\phi
    -\left.\frac{\epsilon\zeta(T+\bar{T})}{2}\sqrt{\frac{3}
{K_{5,Z\bar Z}}}\cos\phi
\right\},  \nonumber\\
m^2&=&m_{3/2}^2
\left\{
          1-3\Bigl[1-\frac{9(S+\bar{S})^2}{\Bigl(3(S+\bar{S})
          +\epsilon\zeta(T+\bar{T})\Bigr)^2}\Bigr]\sin^2\theta \sin^2\phi 
\right.
\nonumber \\
   &-&\Bigl[1-\Bigl(\frac{\epsilon\zeta(T+\bar{T})}
          {(3(S+\bar{S})+\epsilon\zeta(T+\bar{T})}\Bigr)^2\Bigr]
           \cos^2\theta \sin^2\phi \nonumber \\
   &-&\frac{3\epsilon\zeta(T+\bar{T})}{K_{5,Z \bar Z}
          \Bigl(3(S+\bar{S})+\epsilon\zeta(T+\bar{T})\Bigr)}
        \Bigl[2-\frac{\epsilon\zeta(T+\bar{T})}
      {(3(S+\bar{S})+\epsilon\zeta(T+\bar{T})}\Bigr]\cos^2\phi \nonumber \\
   &+&\frac{6\sqrt{3}(S+\bar{S})\epsilon\zeta(T+\bar{T})}
       {\Bigl(3(S+\bar{S})+\epsilon\zeta(T+\bar{T})\Bigr)^2}
           \sin\theta \cos\theta \sin\phi \nonumber \\
   &-&\frac{6\sqrt{3}(S+\bar{S})\epsilon\zeta(T+\bar{T})}
              {\Bigl(3(S+\bar{S})+\epsilon\zeta(T+\bar{T})\Bigr)^2}
            \sqrt{\frac{3}{K_{5,Z \bar Z}}}\sin\theta \sin\phi \cos\phi 
\nonumber \\
   &+&\frac{2\epsilon\zeta(T+\bar{T})}{3(S+\bar{S})+\epsilon\zeta(T+\bar{T})}
            \Bigl[1-\frac{\epsilon\zeta(T+\bar{T})}
         {(3(S+\bar{S})+\epsilon\zeta(T+\bar{T})}\Bigr]\nonumber \\
&\times&
\left.   
        \sqrt{\frac{3}{K_{5,Z \bar Z}}}\cos\theta \sin\phi \cos\phi 
\right\},
\nonumber \\
&{}&  \nonumber\\
A&=&m_{3/2}
\left\{
\sqrt{3}\Bigl[2
       -\frac{9(S+\bar{S})}{3(S+\bar{S})
        +\epsilon\zeta(T+\bar{T})}\Bigr]\sin\theta \sin\phi 
\right.
\nonumber \\
    &-&\frac{3\epsilon\zeta(T+\bar{T})}
          {3(S+\bar{S})+\epsilon\zeta(T+\bar{T})}\cos\theta \sin\phi
\nonumber \\
    &+&
\left.
    \Bigl[K_{5,Z}+\frac{3\epsilon\zeta(T+\bar{T})}
         {3(S+\bar{S})+\epsilon\zeta(T+\bar{T})}\Bigr]
            \sqrt{\frac{3}{K_{5,Z \bar Z}}}\cos\phi    
\right\}, \label{5soft}
\end{eqnarray}
where we have assumed $T=(T+\bar{T})/2$.

In addition, from the gauge kinetic functions (\ref{5gfunc})
we obtain  constraints for values of moduli $S$ and $T$ as,
\begin{eqnarray}
2Re(f^{(1)})=(S+\bar{S})+\epsilon\zeta(T+\bar{T})=2g_{GUT}^{-2}\simeq 4,
\label{5const}
\end{eqnarray}
and 
\begin{eqnarray}
2Re(f^{(2)})=(S+\bar{S})-\frac{3}{2}\epsilon\zeta(T+\bar{T})>0.\label{5const2}
\end{eqnarray}
Here we use eqs.(\ref{asmp2}), (\ref{relazeta}), (\ref{coho}), (\ref{5const})
and (\ref{5const2}), so that we find the
parameter region as $0< \epsilon\zeta(T+\bar{T}) < 8/5$.    
That is almost comparable to the region $0< \alpha(T+\bar{T}) < 2$ 
in the case without 5-branes.

\subsection{Electroweak symmetry breaking and mass spectrum}

Using eq.(\ref{5soft}), we analyze the 5-brane effects on 
electroweak symmetry breaking and the mass spectrum.
Eq.(\ref{5soft}) has total 6 free parameters, that is,
$m_{3/2}$, $\epsilon\zeta(T+\bar{T})$, $\theta$,
$\phi$, and $K_{5,Z \bar Z}$, $K_{5,Z}$. 
Here we use the same notation for $\epsilon\zeta(T+\bar{T})$ as 
the case without 5-brane, i.e., 
$\alpha (T+\bar{T}) \equiv \epsilon\zeta(T+\bar{T})$.
Note that $K_{5,Z \bar Z}$ is positive
definite value due to eq.(\ref{5mgra}), (\ref{asmp1}), 
but $K_{5,Z}$ is any real number.
The parameter $K_{5,Z}$ appears only in the $A$-parameter and 
$K_{5,Z}$-dependence of $A$ is easy to understand, that is, 
$A/\cos \phi$ increases linearly as $K_{5,Z}$ increases.

Now, let us concentrate ourselves to the region with $|F^S|>>|F^T|$, 
that is, $\sin \theta \approx 1$.
In this region,  eq.(\ref{5soft}) is reduced to  
\begin{eqnarray}
M_{1/2}&=&\frac{m_{\frac{3}{2}}}{4}
\left\{
       \sqrt{3}(4-\alpha(T+\bar{T}))\sin\phi
         -\frac{\alpha(T+\bar{T})}{2}\sqrt{\frac{3}{K_{5,Z \bar Z}}}
\cos\phi \right\},  \nonumber \\
m^2&=&m_{3/2}^2
\left\{
          1-3\Bigl[1-\frac{9(4-\alpha(T+\bar{T}))^2}{\Bigl(3(S+\bar{S})
          +\alpha(T+\bar{T})\Bigr)^2}\Bigr]\sin^2\phi 
\right.
\nonumber \\
   &-&\frac{3\alpha(T+\bar{T})}{K_{5,Z \bar Z}
          \Bigl(3(S+\bar{S})+\alpha(T+\bar{T})\Bigr)}
        \Bigl[2-\frac{\alpha(T+\bar{T})}
      {(3(S+\bar{S})+\alpha(T+\bar{T}))}\Bigr]\cos^2\phi \nonumber \\
   &-&\left.
         \frac{6\sqrt{3}(4-\alpha(T+\bar{T}))\alpha(T+\bar{T})}
              {\Bigl(3(S+\bar{S})+\alpha(T+\bar{T})\Bigr)^2}
            \sqrt{\frac{3}{K_{5, Z \bar Z}}}\sin\phi \cos\phi
\right\},
\nonumber \\
&{}&  \nonumber \\
A&=&m_{3/2}
\left\{
\sqrt{3}\Bigl[2
       -\frac{9(4-\alpha(T+\bar{T}))}{3(S+\bar{S})
        +\alpha(T+\bar{T})}\Bigr]\sin\phi 
\right.
\nonumber \\
    &+&
\left.
    \Bigl[K_{5,Z}+\frac{3\alpha(T+\bar{T})}
         {3(S+\bar{S})+\alpha(T+\bar{T})}\Bigr]
            \sqrt{\frac{3}{K_{5,Z \bar Z}}}\cos\phi    
\right\}. \label{dilatonsoft}
\end{eqnarray}

To compare with the results in section 2 without 5-brane, 
here we take $\tan \beta =53$.
Actually, in the case with such large $\tan \beta$ we have stronger 
constraints.
At first, we consider the case with $K_{5,Z \bar Z}=K_{5,Z}=1$.
Fig. 6 shows $M_{1/2}$, $m$ and $A$ for $\phi =\pi/4$ 
as well as $m_{3/2}=2.0$ TeV, $\theta =\pi/2$,
and $K_{5,Z\bar Z}=K_{5,Z}=1$.
Obviously, the ratios, $A/M_{1/2}$ and $m/M_{1/2}$, are 
different from eq.(\ref{sumrule}). 
The $A$-parameter is always positive when $\phi =\pi/4$.
For small $\alpha(T+\bar{T})$, the dominant terms are the first
term of $M_{1/2}$, and the second term of $m^2$.
The 5-brane SUSY breaking effect $F^Z$ within $0<\sin\phi<1$ 
reduces $M_{1/2}$, but increases $m^2$.
That is obvious from the comparison between Fig. 6 
and eq.(\ref{sumrule}).
Therefore we obtain 
$m_{\tilde{\tau}_1}^2>m_{\tilde{\chi}_1^0}^2 \sim M_1^2$ at
small $\alpha(T+\bar{T})$ and the lightest
neutralino is almost Bino-like.
When $\alpha(T+\bar{T})$ become larger, the third and forth 
term of $m^2$ become more sizable and $m^2$ itself is decreased.
Thus the excluded region by constraint $m_{\tilde{\tau}_1}^2>0$ is larger
than the case without the 5-brane effect. 
Such behavior on electroweak symmetry breaking and the constraint 
$m_{\tilde{\tau}_1}^2>0$ is shown in Fig. 7a  
for $\theta =\pi/2$, $\phi =\pi/4$ and $K_{5,Z \bar Z}=K_{5,Z}=1$.
In the allowed region, the lightest neutralino is almost Bino-like.
Fig. 7b shows the case with $\phi =\pi/5$.
In the 5-brane dominant case, $\cos \phi \to 1$, most of the region 
is excluded.
Figs. 8a and 8b show $\alpha(T+\bar T)$-dependence and $\phi$-dependence 
of representative superparticle masses, e.g. 
the lightest neutralino, chargino, stop and stau.
The $\phi$-dependence is rather larger than 
$\alpha(T+\bar T)$-dependence.
When we vary $\phi$, the lightest Higgs mass change 
$132$ GeV $< m_{h^0} < 138$ GeV in the allowed region for 
$m_{3/2}= 1$ TeV and $\theta =\pi/2$.
The lightest Higgs mass is rather stable for change of 
$\alpha(T+\bar T)$.

Next, we calculate $K_{5,Z \bar Z}$ contribution fixing 
$K_{5,Z} =1$.
Fig. 9 shows $M_{1/2}$, $m$ and $|A|$ for $m_{3/2}=2.0$ TeV, 
$\theta =\pi/2$, $\phi =\pi/4$, $2\alpha Re(T)=1.0$
and $K_{5,Z}=1$.
Obviously, the $A/M_{1/2}$ and $m/M_{1/2}$, are 
different from eq.(\ref{sumrule}), again. 
The $A$-parameter changes from positive to negative 
around $K_{5,Z \bar Z} =2.5$ for $\phi =\pi/4$ and $2\alpha Re(T)=1.0$.
The terms of
the 5-brane contribution of $M_{1/2}$ and $m^2$ are proportional to
$\sqrt{3/K_{5,Z \bar Z}}$ or $3/K_{5,Z \bar Z}$.
When
$K_{5,Z \bar Z}$ increases, then these terms are suppressed and these soft
parameters become close to those without 5-brane except for 
the existence of
the ``reducing'' factor $\sin\phi$.
Therefore  the effects of $\sin\phi$ is favorable for the
$m_{\tilde{\tau}_1}^2>0$ condition 
and it leads to $m_{\tilde{\tau}_1}^2>m_{\tilde{\chi}_1^0}^2$. 
Fig. 10 shows such $K_{5,Z \bar Z}$-dependence of the regions excluded by 
the electroweak symmetry breaking and the constraint 
$m_{\tilde{\tau}_1}^2>0$ as well as the stau LSP region.
Actually, large $K_{5,Z \bar Z}$ is useful to increase 
$m_{\tilde{\tau}_1}^2$, but it is not favorable for 
successful electroweak symmetry breaking.

Finally, we show $K_{5,Z}$-dependence, which has the linear effect 
only on $A$.
Figs. 11 show the $K_{5,Z}$ effect on the electroweak symmetry breaking,  
the constraint $m_{\tilde{\tau}_1}^2>0$ and the LSP 
for $m_{3/2}=2.0$ TeV, 
$\phi = \pi/4$ and $K_{5,Z \bar Z}=1$.
As $K_{5,Z}$ increases, the electroweak symmetry breaking condition 
is relaxed, but the constraint $m_{\tilde{\tau}_1}^2>0$ becomes 
severe.
Actually, the region with unsuccessful electroweak symmetry breaking 
disappears at $K_{5,Z}>1.4$, but  at $K_{5,Z}>3.2$ 
all the region is excluded because of $m_{\tilde{\tau}_1}^2 <0$.
Fig. 12 shows $K_{5,Z}$-dependence of representative superparticle 
masses.
The lightest Higgs mass is rather stable against change of $K_{5,Z}$.

\section{Conclusions}

We have studied soft SUSY breaking parameters in M-theory 
with and without 5-brane moduli effects.
We investigate successful electroweak symmetry breaking and 
the constraint $m_{\tilde{\tau}_1}^2>0$.
In the allowed regions, we have shown mass spectra.

We have considered the large $\tan \beta$ scenario, 
because these constraints are severe for large $\tan \beta$.
In the case without 5-brane effects, 
the electroweak symmetry breaking conditions are relaxed in the 
strong coupling region $\alpha(T +\bar T) > 1.5$.
In such region, however, the constraint $m_{\tilde{\tau}_1}^2>0$ 
becomes strong in particular for large $\theta$.
In any allowed region, the LSP is the stau.

The positive $D$-term contribution $m_{D2}^2$ can relax 
the constraint $m_{\tilde{\tau}_1}^2>0$ and lead to 
the region with the neutralino LSP.
The negative $D$-term contribution $m_{D1}^2$ can relax 
the electroweak symmetry breaking conditions.
In the suitable value of $m_{D1}^2$ we have the widest region 
with the neutralino LSP.

For small $\tan \beta$, in particular $\tan \beta < 50$, 
the electroweak symmetry breaking is realized easily.
However, for large $\alpha(T +\bar T)$ and 
large $\theta$, the region with $m_{\tilde{\tau}_1}^2 <0$ and the region 
with the stau LSP remain.

The 5-brane effect can relax the electroweak symmetry breaking condition, 
but it makes the $m_{\tilde{\tau}_1}^2>0$ constraint severe.

Soft terms in M-theory with 5-branes include more free parameters.
Thus, we will leave to future work systematic study of the 
whole parameter space on other phenomenological aspects, 
e.g. cosmological implications and the $b \to s \gamma$ decay \cite{bs}, 
which leads to anther constraint for large $\tan \beta$ and lighter 
superparticles in the universal case.

\section*{Acknowledgement} 
The authors would like to thank K.~Enqvist and S.-J. Rey 
for useful discussions.
This work was partially supported by the Academy of Finland (no. 44129).

\newpage
\section*{Figure Captions}

\renewcommand{\labelenumi}{Fig.~\arabic{enumi}}

\begin{enumerate}

\item The ratios $-A/M_{1/2}$ and $m/M_{1/2}$ against $2\alpha Re(T)$.
The three solid curves correspond to $-A/M_{1/2}$ for 
$\theta =0, \pi/5$ and $\pi/2$.
The two dotted curves correspond to $m/M_{1/2}$ for 
$\theta =\pi/5$ and $\pi/2$.

\item The excluded regions by the
electroweak breaking condition and the constraint $m_{\tilde \tau_1}^2>0$. 
Also the regions with $m_{\tilde \tau_1}^2 > m^2_{\tilde \chi_1}$ are 
shown.
Fig.2a, 2b, 2c and 2d correspond to the cases with 
$\theta$=$\pi/5$, $\pi/4$, $\pi/2$, $3\pi/4$. 

\item Effects of the $D$-term contribution $m_{D2}^2$ 
      on the excluded region for
      $\theta$=$\pi/2$. 
      Figs.3a and 3b correspond to the cases with 
      $(m_{D1}^2,m_{D2}^2)=(0,0.5)$ and $(0,1.0)$ [(TeV)$^2$], 
      respectively. 

\item Effects of the $D$-term contribution $m_{D1}^2$ 
      on the excluded region for
      $\theta$=$\pi/2$. 
      Figs. 4a, 4b and 4c correspond to the cases with 
      $(m_{D1}^2,m_{D2}^2)=(-0.05,0)$, $(-2.0,0)$ and $(-5.0,0)$ [(TeV)$^2$], 
      respectively.

\item The excluded region for tan$\beta$.  
      Fig. 5a corresponds to the case with $\theta$=$\pi/5$ and 
      $m_{3/2}$=1.0 TeV for $M_{top}$=175.0 GeV.
      Similarly, Fig. 5b corresponds to the case with 
      $\theta$=$\pi/2$ and  $m_{3/2}$=2.0 TeV
      for $M_{top}$=175.0 GeV. 

\item The soft parameters, $M_{1/2}$, $|A|$ and $m$ in the case with 
the 5-brane effect for $m_{3/2}=2.0$ TeV, $\theta =\pi/2$, $\phi =\pi/4$ 
and $K_{5,Z\bar Z}=K_{5,Z}=1$.

\item 5-brane effects on the electroweak symmetry breaking,  
the constraint $m_{\tilde{\tau}_1}^2>0$ and the LSP
for $\theta =\pi/2$ and 
$K_{5,Z\bar Z}=K_{5,Z}=1$.
Figs. 7a and 7b correspond to $\phi=\pi/4$ and $\pi/5$.

\item The superparticle spectra against $2\alpha Re(T)$ and 
$\pi$.
Fig. 8a shows $2\alpha Re(T)$-dependence of
the lightest neutralino, chargino, stau and stop masses for 
$m_{3/2}=1.0$ TeV, $\theta = \pi/2$, $\pi/4$ and $K_{5,Z\bar Z}=K_{5,Z}=1$.
Similarly, Fig. 8b shows $\phi$-dependence of the superparticel masses for 
$m_{3/2}=1.0$ TeV, $\theta = \pi/2$, $2\alpha Re(T)=0.5$ and 
$K_{5,Z\bar Z}=K_{5,Z}=1$.

\item The soft parameters, $M_{1/2}$, $|A|$ and $m$ 
when we vary $K_{5,Z\bar Z}$.
The other parameters are fixed 
$m_{3/2}=2.0$ TeV, $\theta =\pi/2$, $\phi =\pi/4$, $2\alpha Re(T)=1.0$
and $K_{5,Z}=1$.

\item $K_{5,Z\bar Z}$-dependence of the electroweak symmetry breaking,  
the constraint $m_{\tilde{\tau}_1}^2>0$ and the LSP for 
$m_{3/2}=2.0$ TeV, $\theta =\pi/2$, $\phi =\pi/4$ 
and $K_{5,Z}=1$.

\item $K_{5,Z}$-dependence of the electroweak symmetry breaking,  
the constraint $m_{\tilde{\tau}_1}^2>0$ and the LSP for 
$m_{3/2}=2.0$ TeV, $\theta =\pi/2$, $\phi =\pi/4$ 
and $K_{5,Z \bar Z}=1$.
Fig. 11a and 11b correspond to the regions with 
$-1.0 \leq K_{5,Z} \leq 1.0$ and $1.0 \leq K_{5,Z} \leq 2.0$, 
respectively.

\item The superparticle spectrum against $K_{5,Z}$ for 
$m_{3/2}=1.0$ TeV, $\theta = \pi/2$, $\pi/4$, 
$2\alpha Re(T)=0.5$ and $K_{5,Z\bar Z}=1$.

\end{enumerate}

%\newpage
%\section*{Ordering of Figs.}

 ${}\vspace{-192pt}$

 \begin{figure}[h]     

 ${}\vspace{-80pt}$

 \begin{minipage}{80mm}
  \epsfxsize=110mm     
  \begin{center}
   \leavevmode
   \epsfbox{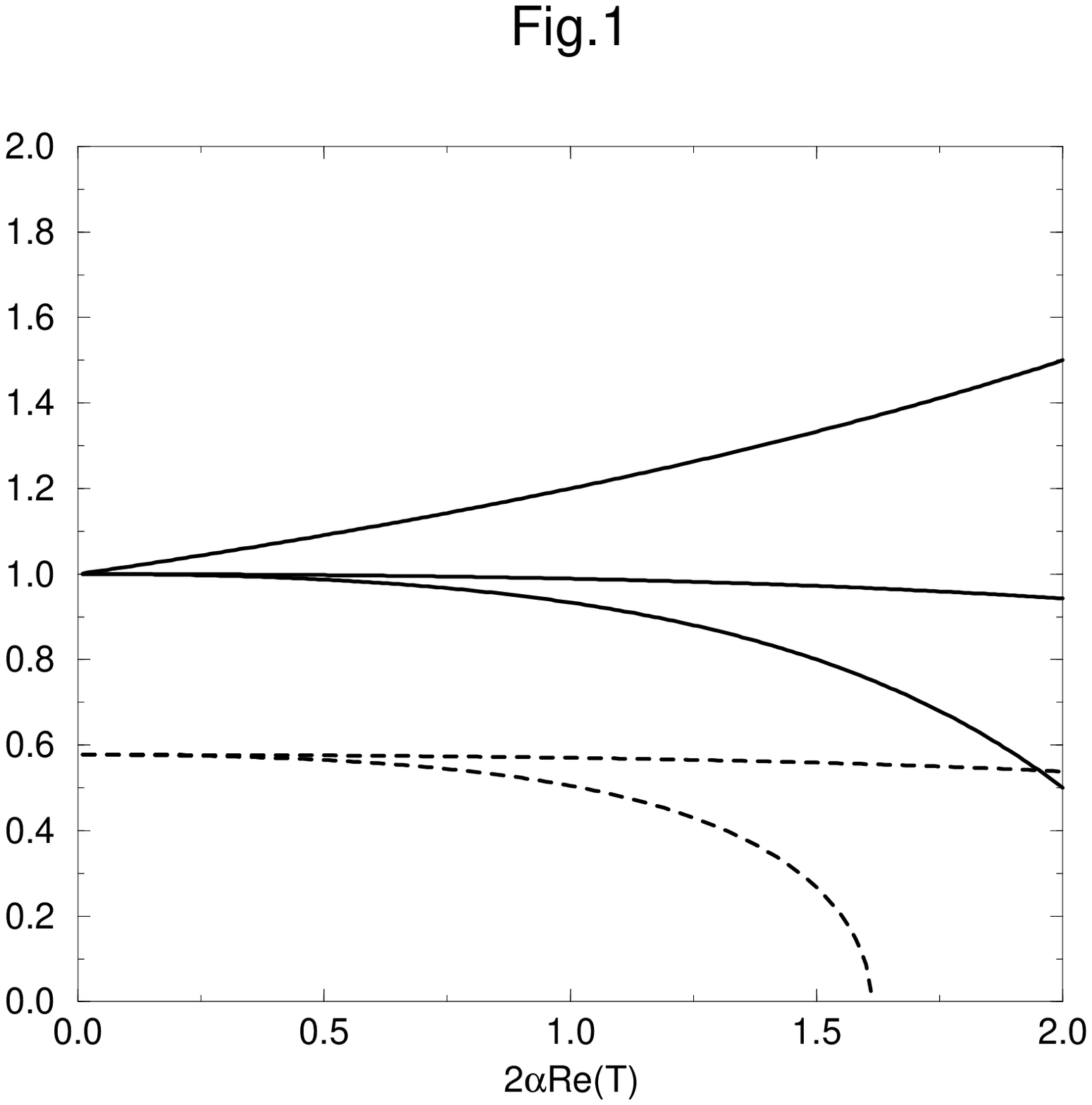}
  \end{center}
 % \caption{}       
 % \label{}       
 \end{minipage}

 ${}\vspace{-122pt}$

 \begin{minipage}{80mm}
  \epsfxsize=110mm  
  \begin{center}
   \leavevmode
   \epsfbox{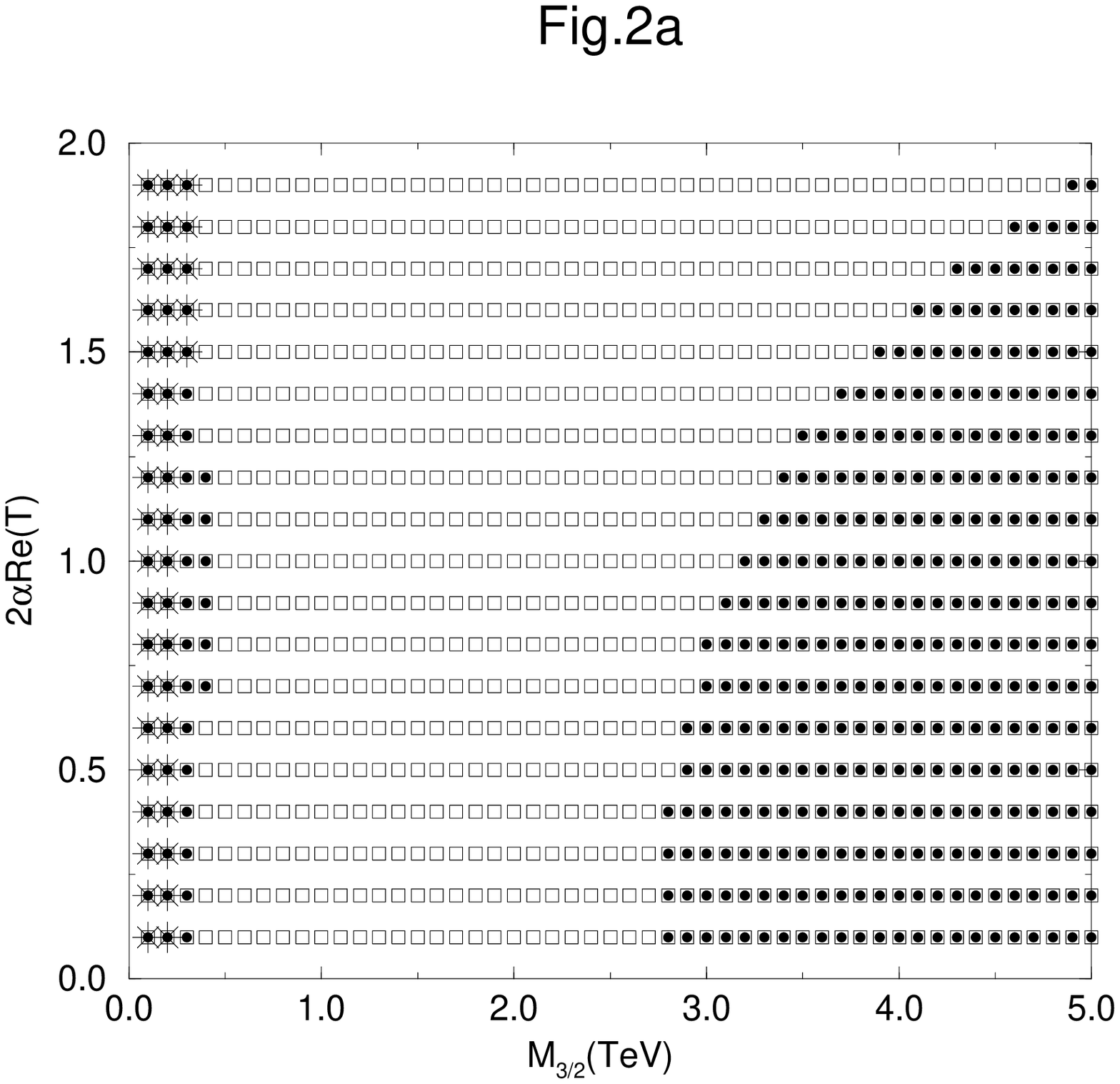}
  \end{center}
 % \caption{}       
 % \label{}       
 \end{minipage}
 \end{figure}

%%%%%%%%%%%%%%%%%%%%%%%%%%%%%%%%%%%%%%%%%%%%%%%%%%%%%%%%%%%%%%%%%%%%%%%%%

 ${}\vspace{-192pt}$

 \begin{figure}[h]     

 ${}\vspace{-80pt}$

 \begin{minipage}{80mm}
  \epsfxsize=110mm     
  \begin{center}
   \leavevmode
   \epsfbox{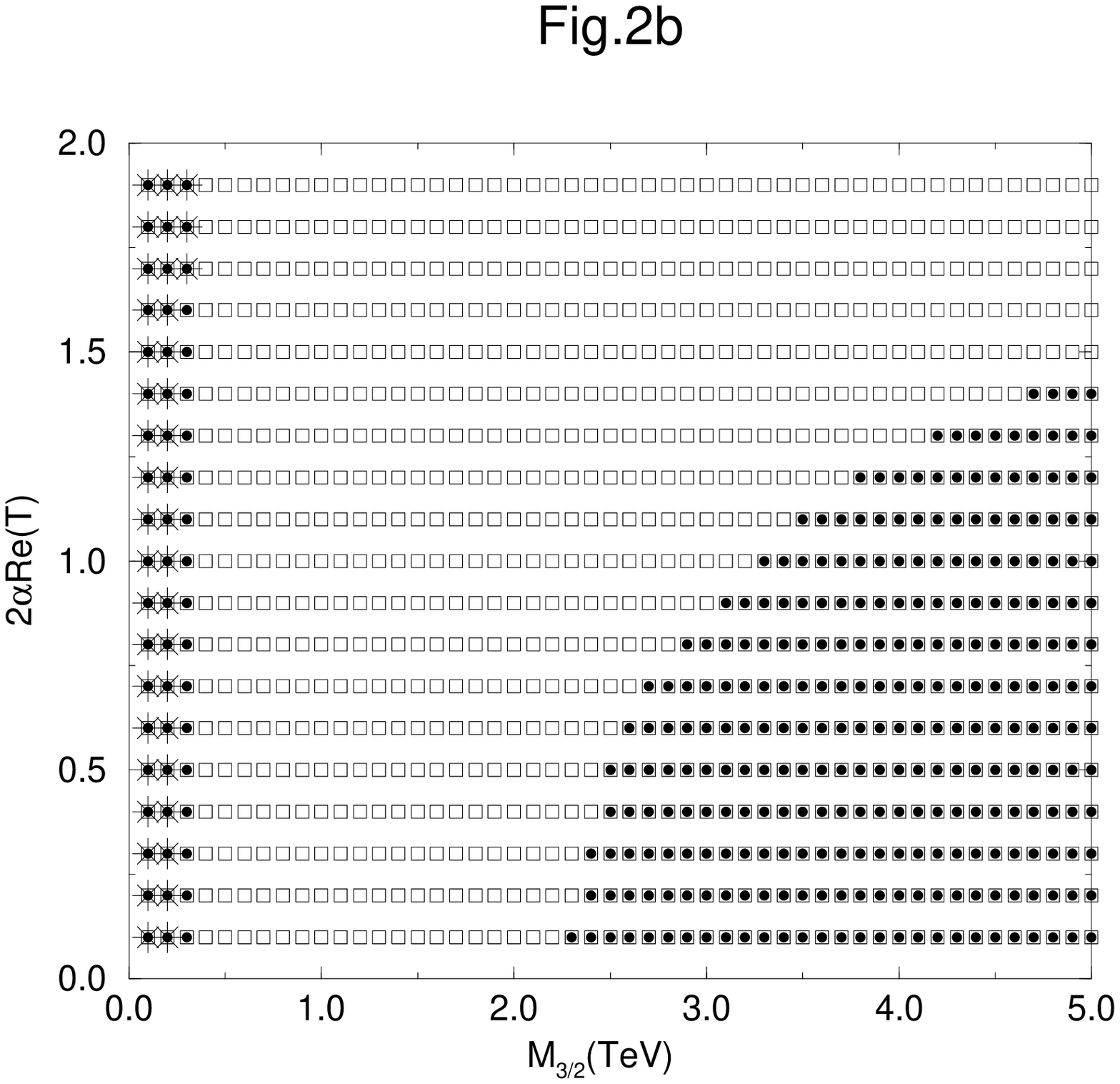}
  \end{center}
 % \caption{}       
 % \label{}       
 \end{minipage}

 ${}\vspace{-122pt}$

 \begin{minipage}{80mm}
  \epsfxsize=110mm  
  \begin{center}
   \leavevmode
   \epsfbox{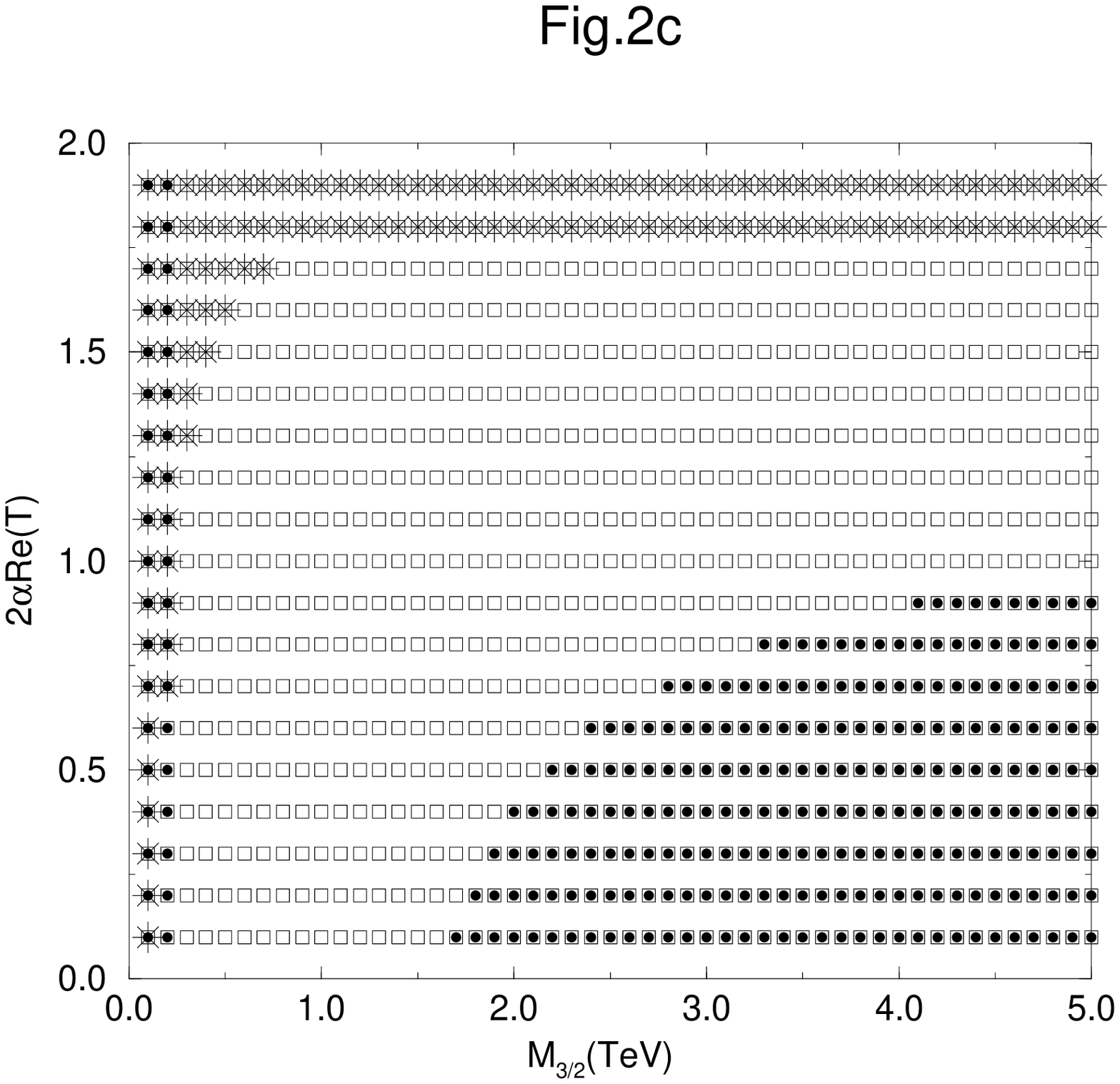}
  \end{center}
 % \caption{}       
 % \label{}       
 \end{minipage}
 \end{figure}

%%%%%%%%%%%%%%%%%%%%%%%%%%%%%%%%%%%%%%%%%%%%%%%%%%%%%%%%%%%%%%%%%%%%%%%%%

 ${}\vspace{-192pt}$

 \begin{figure}[h]     

 ${}\vspace{-80pt}$

 \begin{minipage}{80mm}
  \epsfxsize=110mm     
  \begin{center}
   \leavevmode
   \epsfbox{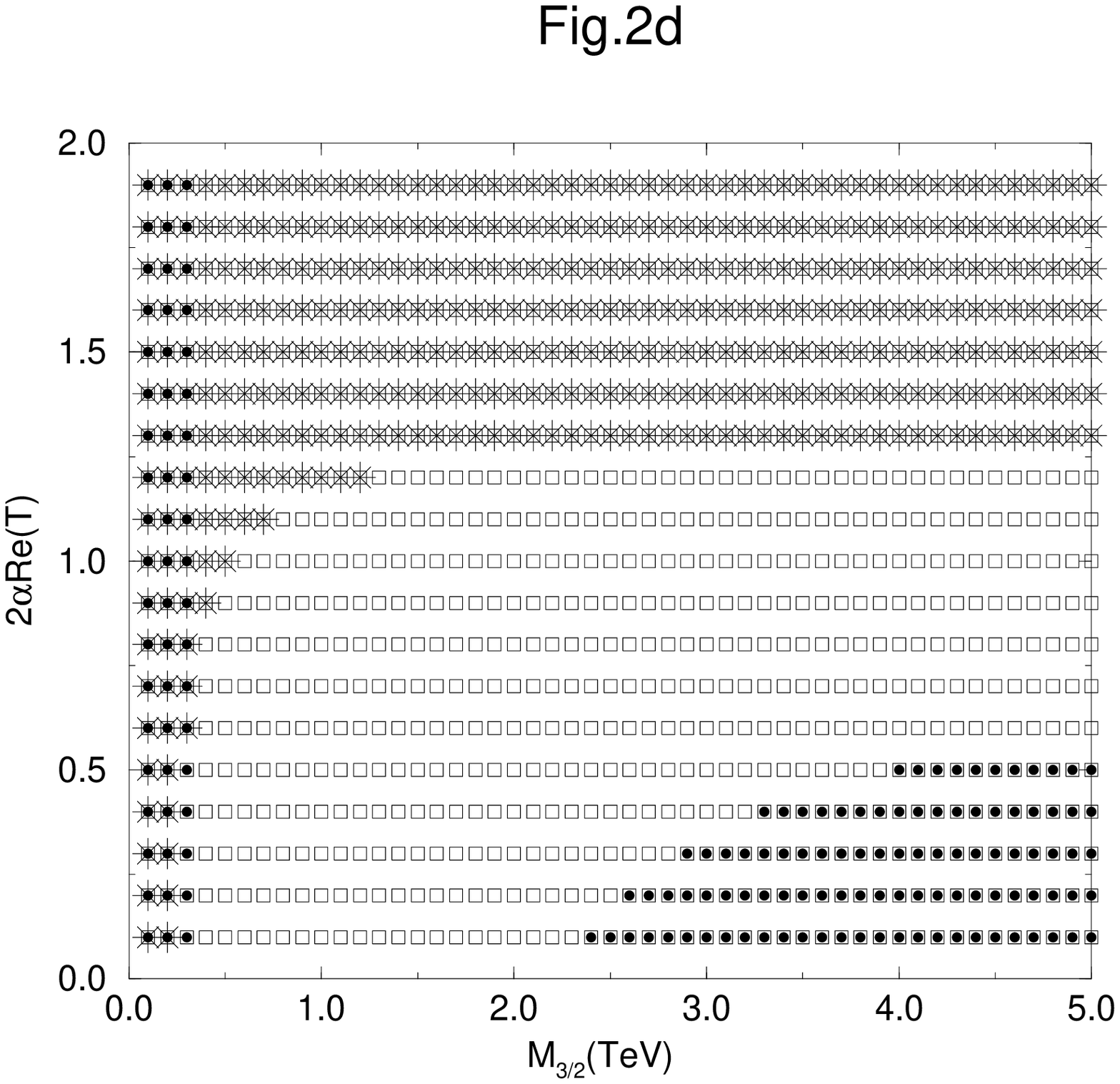}
  \end{center}
 % \caption{}       
 % \label{}       
 \end{minipage}

 ${}\vspace{-122pt}$

 \begin{minipage}{80mm}
  \epsfxsize=110mm  
  \begin{center}
   \leavevmode
   \epsfbox{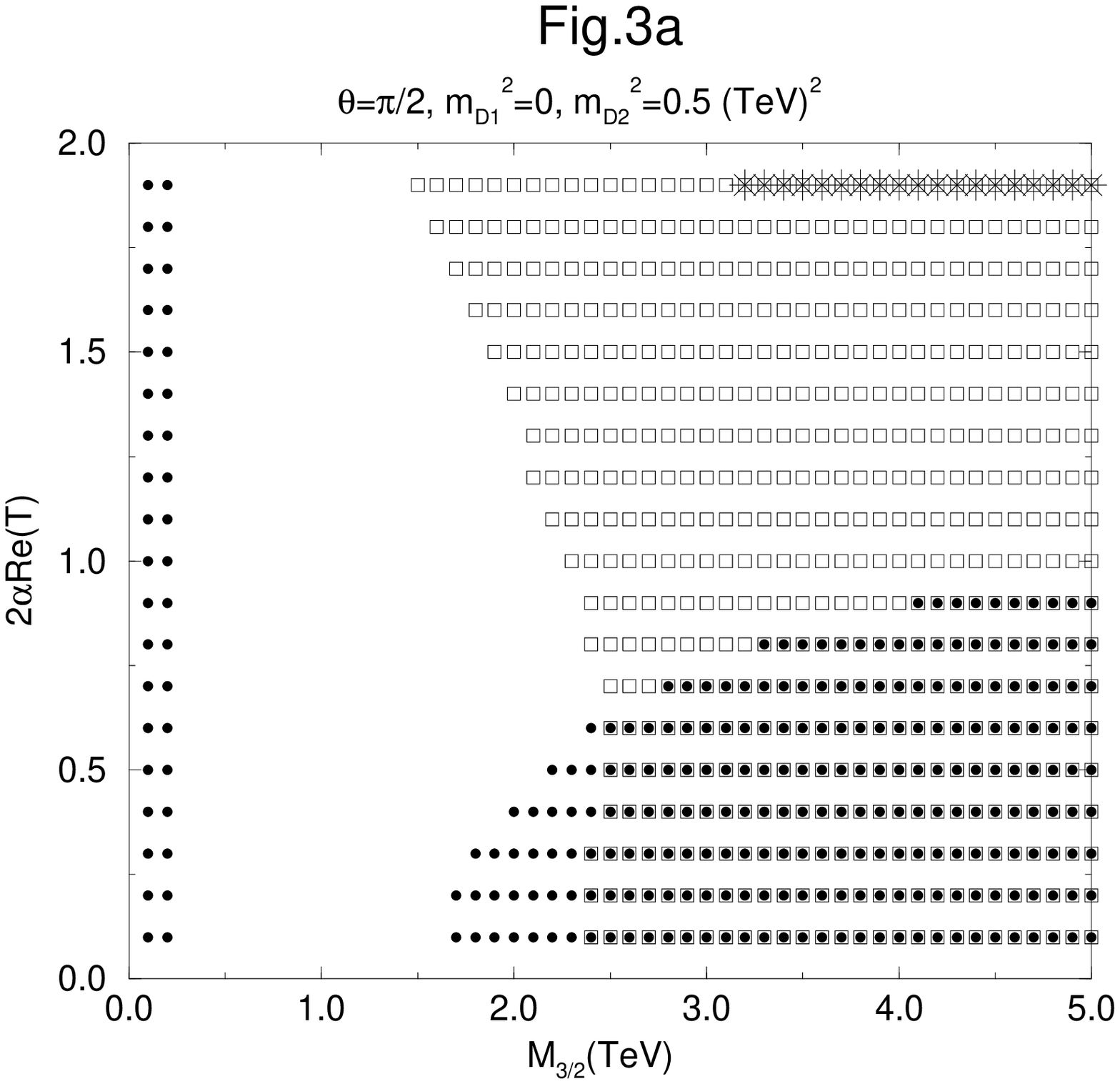}
  \end{center}
 % \caption{}       
 % \label{}       
 \end{minipage}
 \end{figure}

%%%%%%%%%%%%%%%%%%%%%%%%%%%%%%%%%%%%%%%%%%%%%%%%%%%%%%%%%%%%%%%%%%%%%%%%%

 ${}\vspace{-192pt}$

 \begin{figure}[h]     

 ${}\vspace{-80pt}$

 \begin{minipage}{80mm}
  \epsfxsize=110mm     
  \begin{center}
   \leavevmode
   \epsfbox{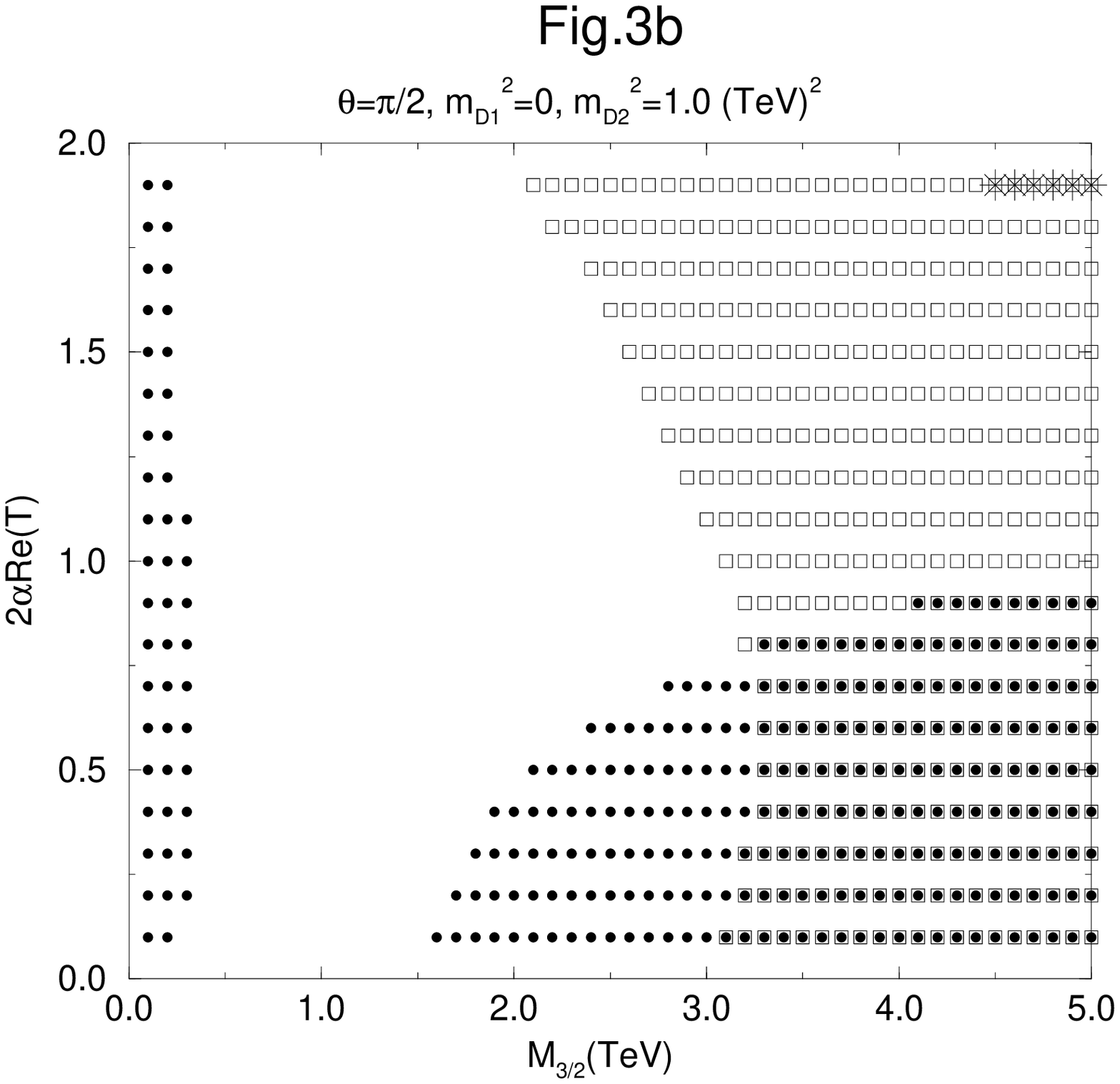}
  \end{center}
 % \caption{}       
 % \label{}       
 \end{minipage}

 ${}\vspace{-122pt}$

 \begin{minipage}{80mm}
  \epsfxsize=110mm  
  \begin{center}
   \leavevmode
   \epsfbox{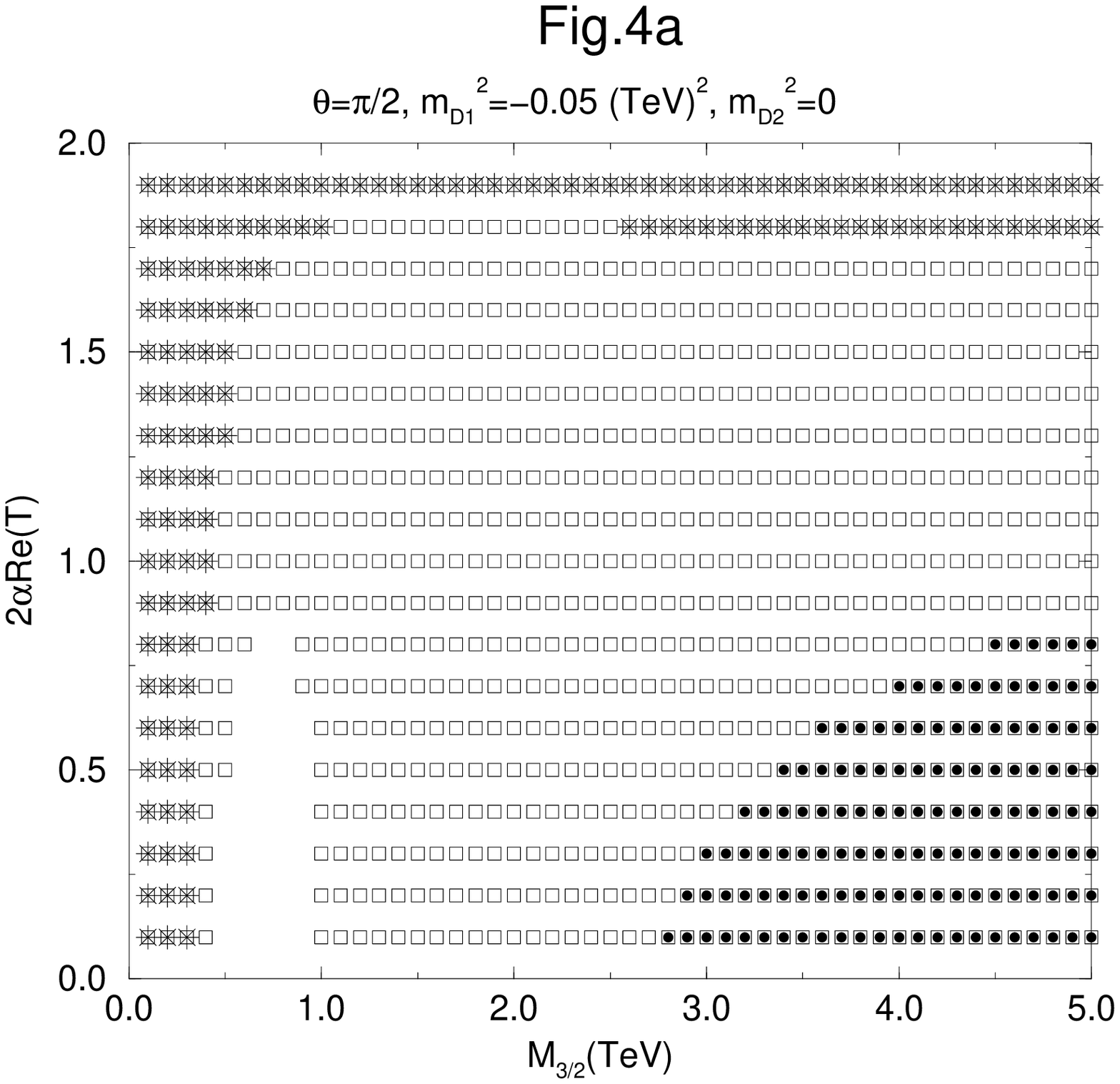}
  \end{center}
 % \caption{}       
 % \label{}       
 \end{minipage}
 \end{figure}

%%%%%%%%%%%%%%%%%%%%%%%%%%%%%%%%%%%%%%%%%%%%%%%%%%%%%%%%%%%%%%%%%%%%%%%%%

 ${}\vspace{-192pt}$

 \begin{figure}[h]     

 ${}\vspace{-80pt}$

 \begin{minipage}{80mm}
  \epsfxsize=110mm     
  \begin{center}
   \leavevmode
   \epsfbox{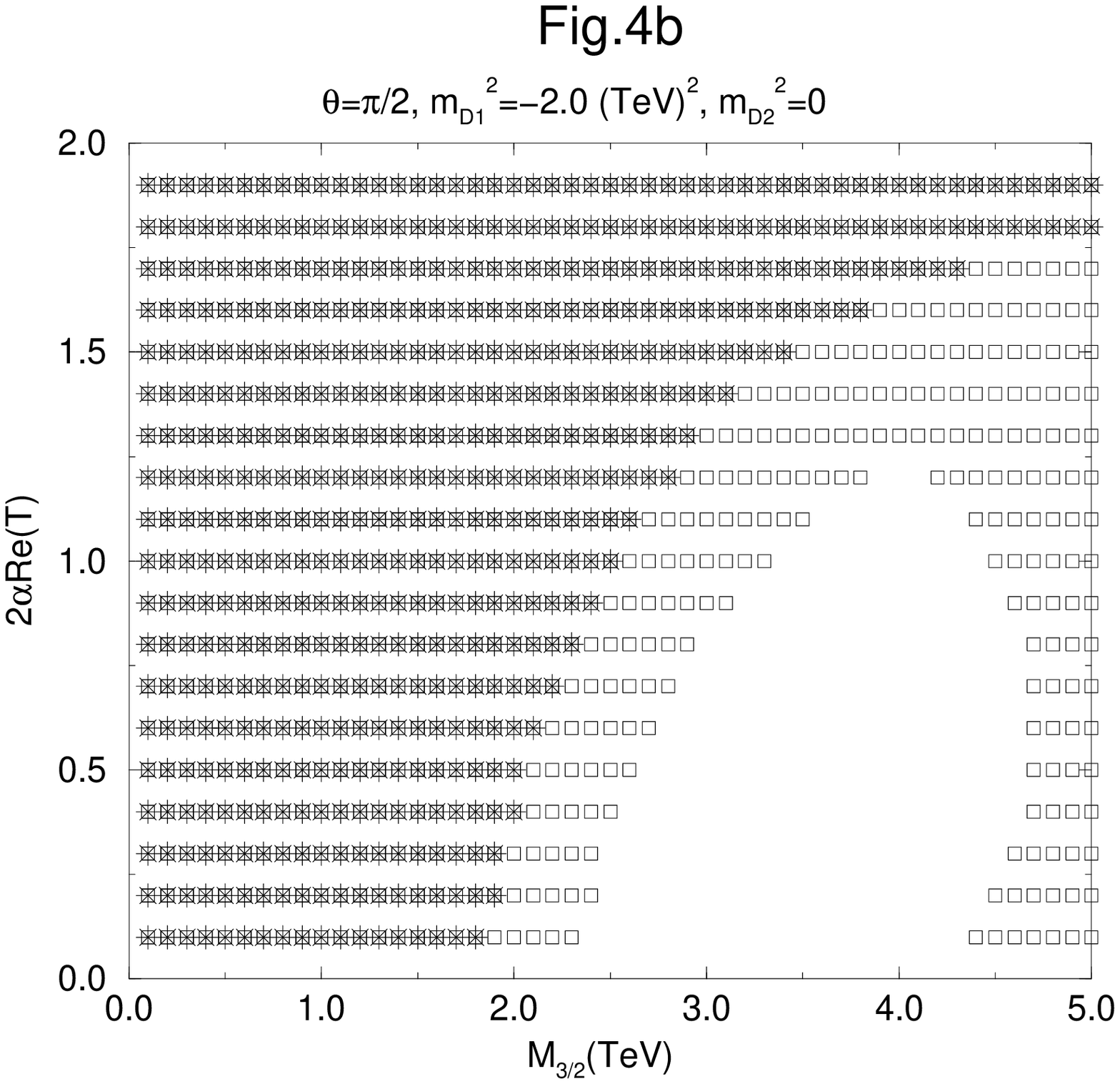}
  \end{center}
 % \caption{}       
 % \label{}       
 \end{minipage}

 ${}\vspace{-122pt}$

 \begin{minipage}{80mm}
  \epsfxsize=110mm  
  \begin{center}
   \leavevmode
   \epsfbox{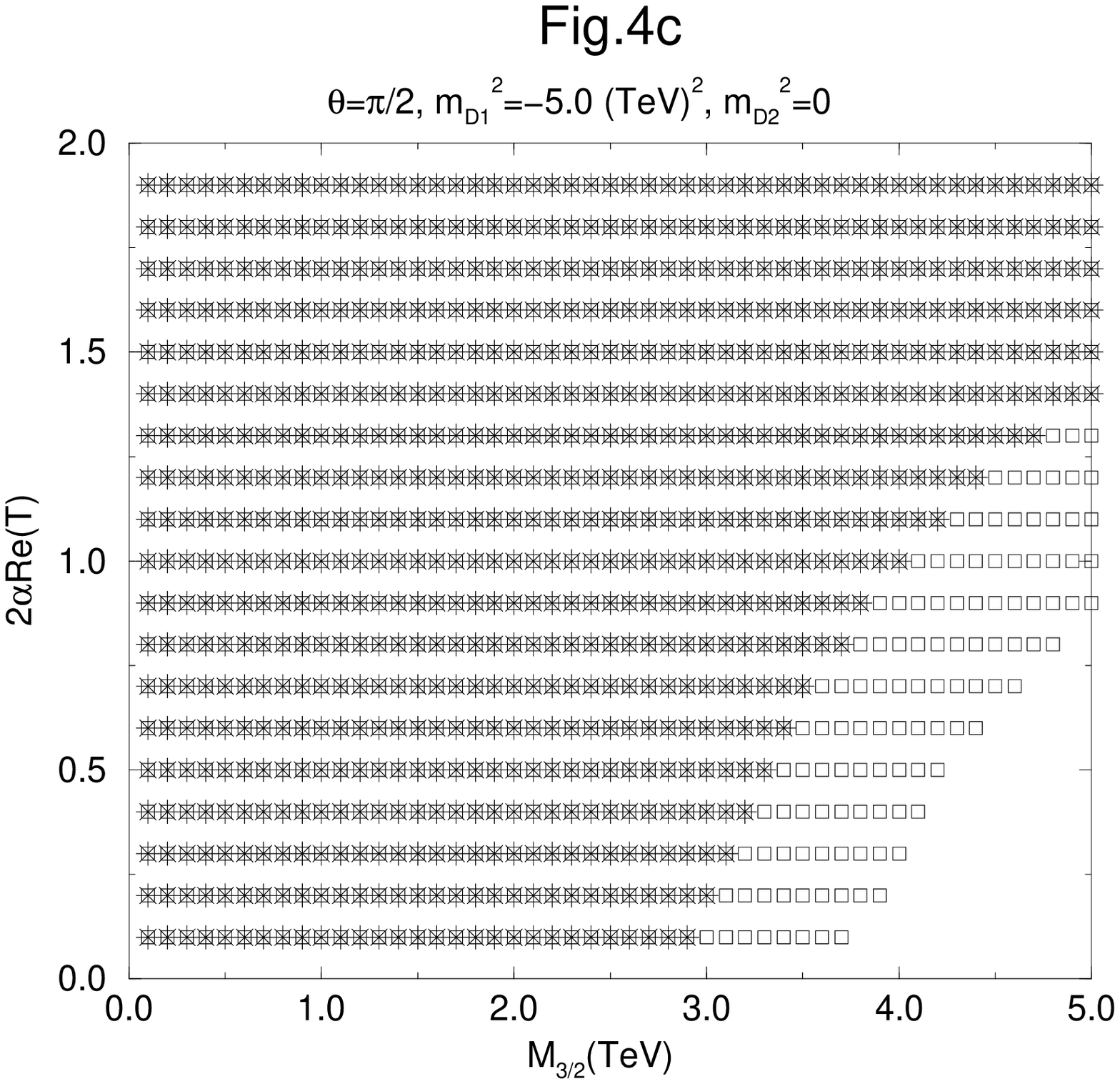}
  \end{center}
 % \caption{}       
 % \label{}       
 \end{minipage}
 \end{figure}

%%%%%%%%%%%%%%%%%%%%%%%%%%%%%%%%%%%%%%%%%%%%%%%%%%%%%%%%%%%%%%%%%%%%%%%%%

 ${}\vspace{-192pt}$

 \begin{figure}[h]     

 ${}\vspace{-80pt}$

 \begin{minipage}{80mm}
  \epsfxsize=110mm     
  \begin{center}
   \leavevmode
   \epsfbox{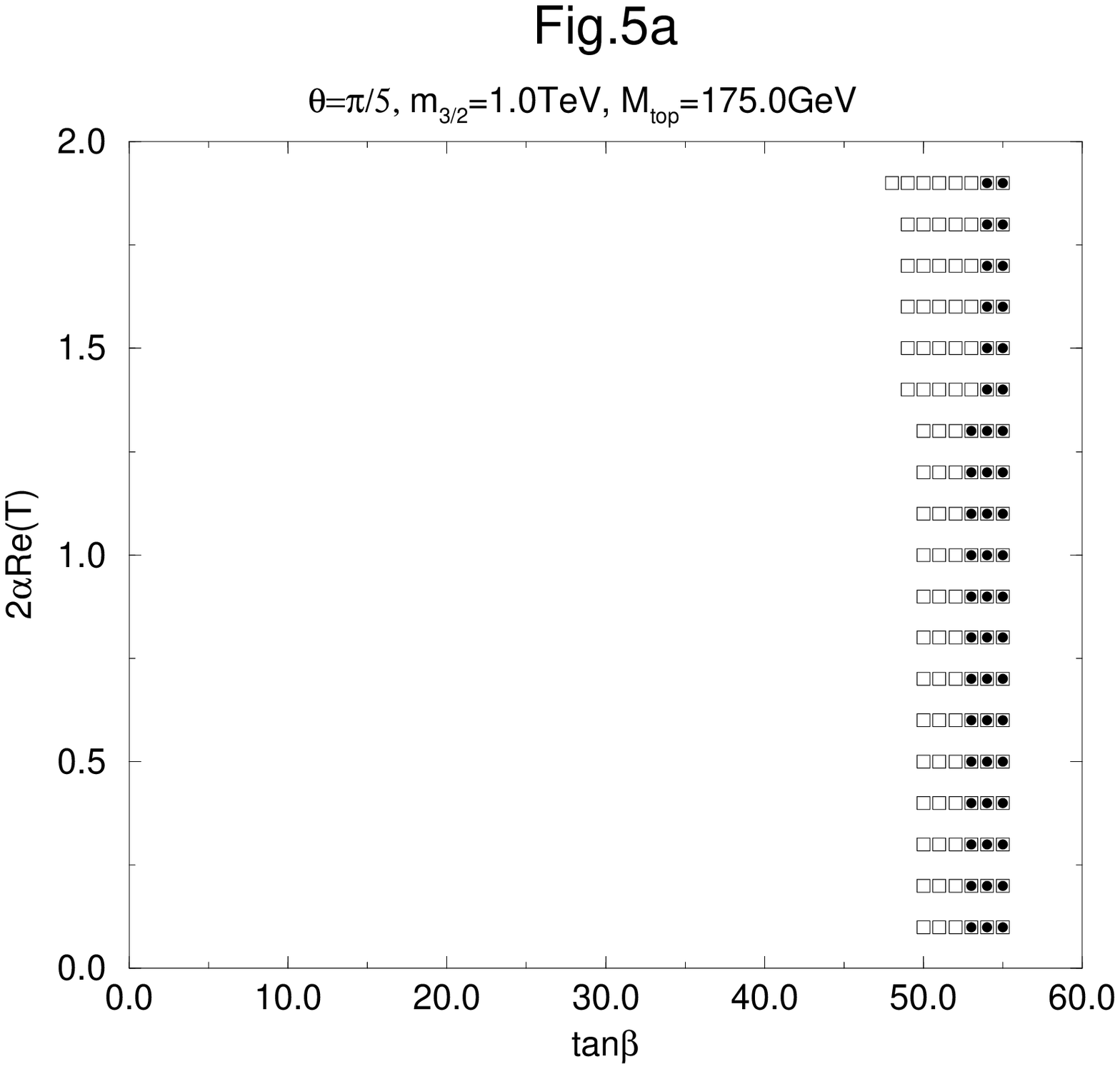}
  \end{center}
 % \caption{}       
 % \label{}       
 \end{minipage}

 ${}\vspace{-122pt}$

 \begin{minipage}{80mm}
  \epsfxsize=110mm  
  \begin{center}
   \leavevmode
   \epsfbox{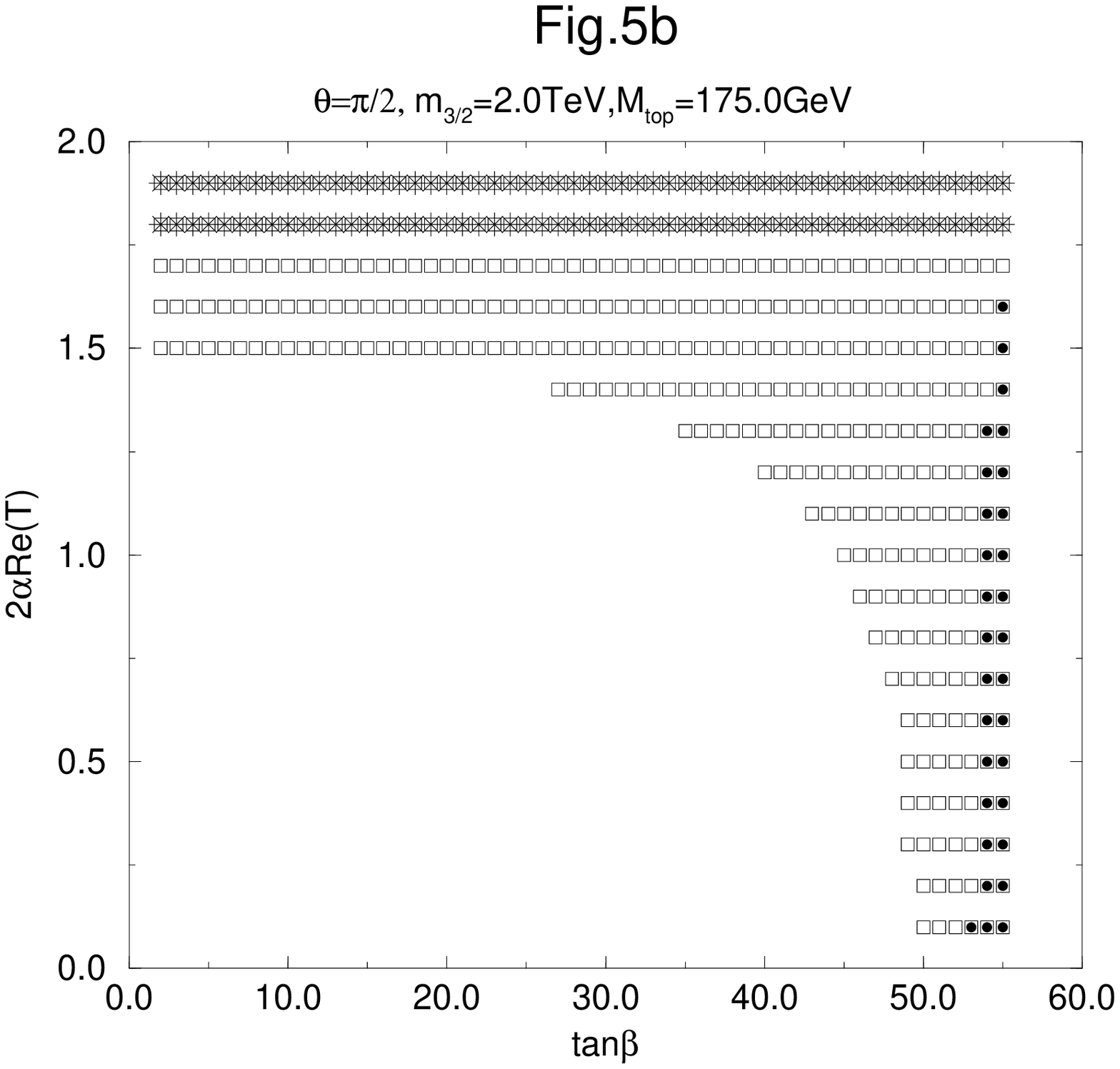}
  \end{center}
 % \caption{}       
 % \label{}       
 \end{minipage}
 \end{figure}

%%%%%%%%%%%%%%%%%%%%%%%%%%%%%%%%%%%%%%%%%%%%%%%%%%%%%%%%%%%%%%%%%%%%%%%%%

 ${}\vspace{-192pt}$

 \begin{figure}[h]     

 ${}\vspace{-80pt}$

 \begin{minipage}{80mm}
  \epsfxsize=110mm     
  \begin{center}
   \leavevmode
   \epsfbox{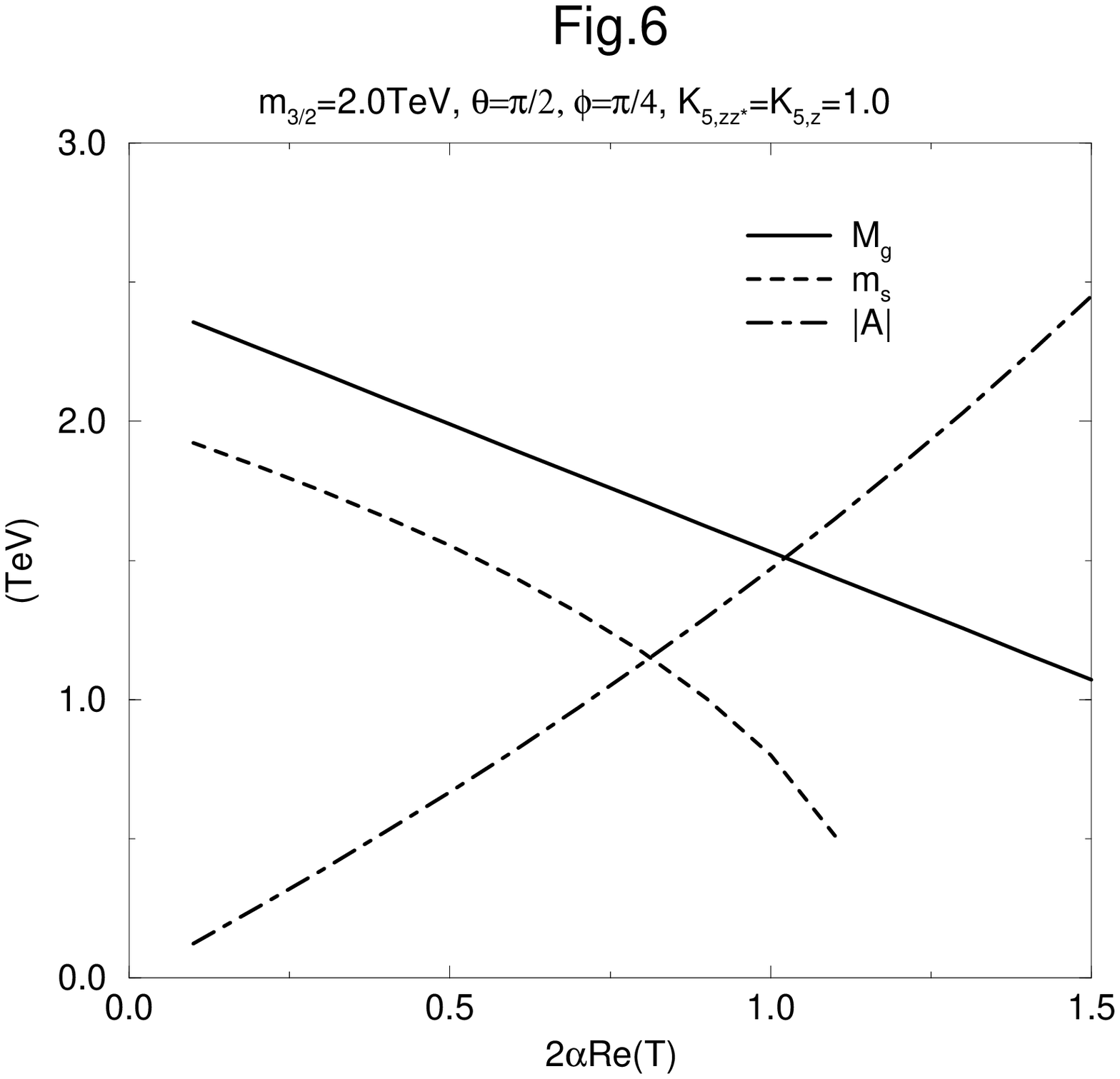}
  \end{center}
 % \caption{}       
 % \label{}       
 \end{minipage}

 ${}\vspace{-122pt}$

 \begin{minipage}{80mm}
  \epsfxsize=110mm  
  \begin{center}
   \leavevmode
   \epsfbox{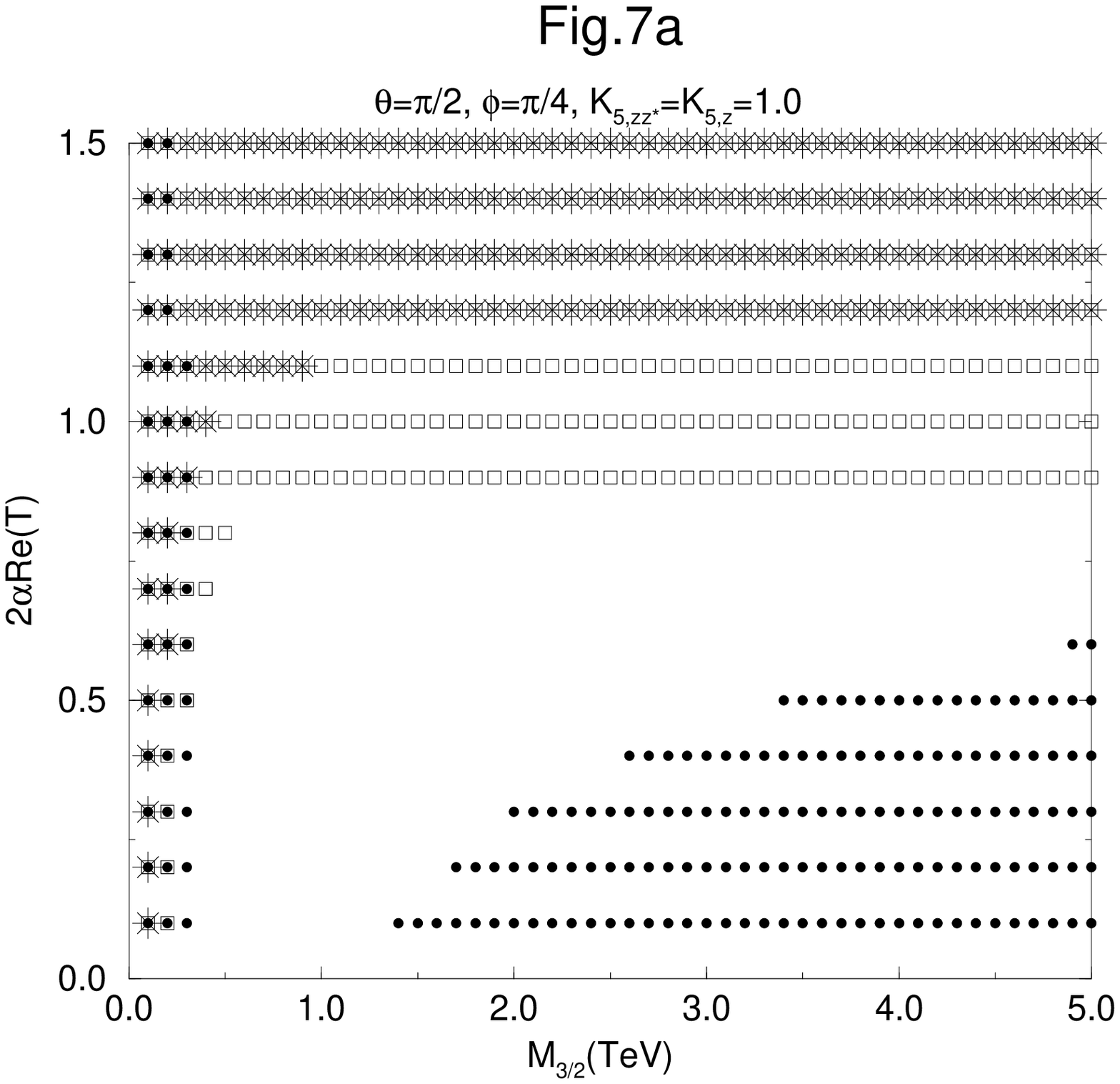}
  \end{center}
 % \caption{}       
 % \label{}       
 \end{minipage}
 \end{figure}

%%%%%%%%%%%%%%%%%%%%%%%%%%%%%%%%%%%%%%%%%%%%%%%%%%%%%%%%%%%%%%%%%%%%%%%%%

 ${}\vspace{-192pt}$

 \begin{figure}[h]     

 ${}\vspace{-80pt}$

 \begin{minipage}{80mm}
  \epsfxsize=110mm     
  \begin{center}
   \leavevmode
   \epsfbox{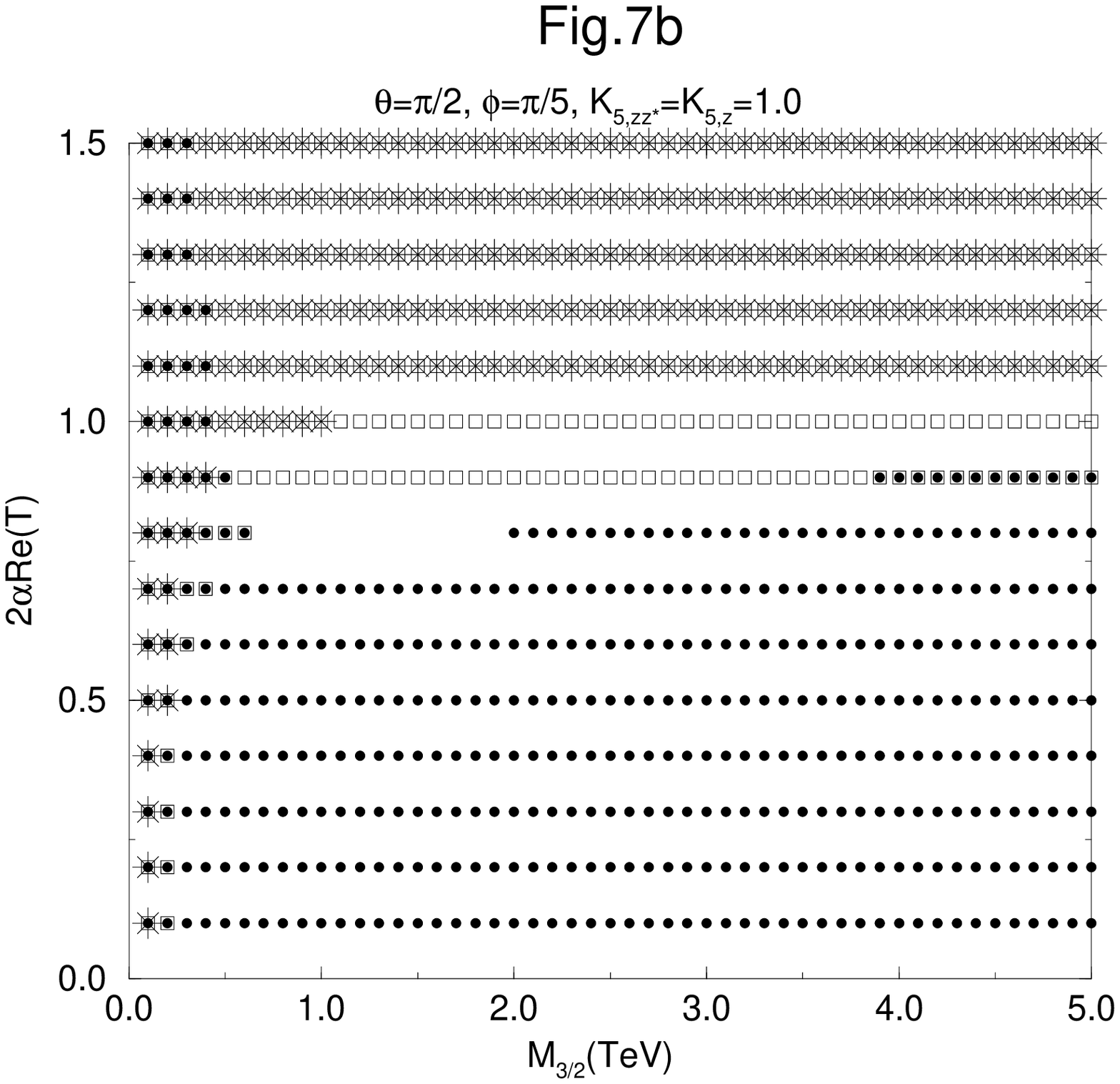}
  \end{center}
 % \caption{}       
 % \label{}       
 \end{minipage}

 ${}\vspace{-122pt}$

 \begin{minipage}{80mm}
  \epsfxsize=110mm  
  \begin{center}
   \leavevmode
   \epsfbox{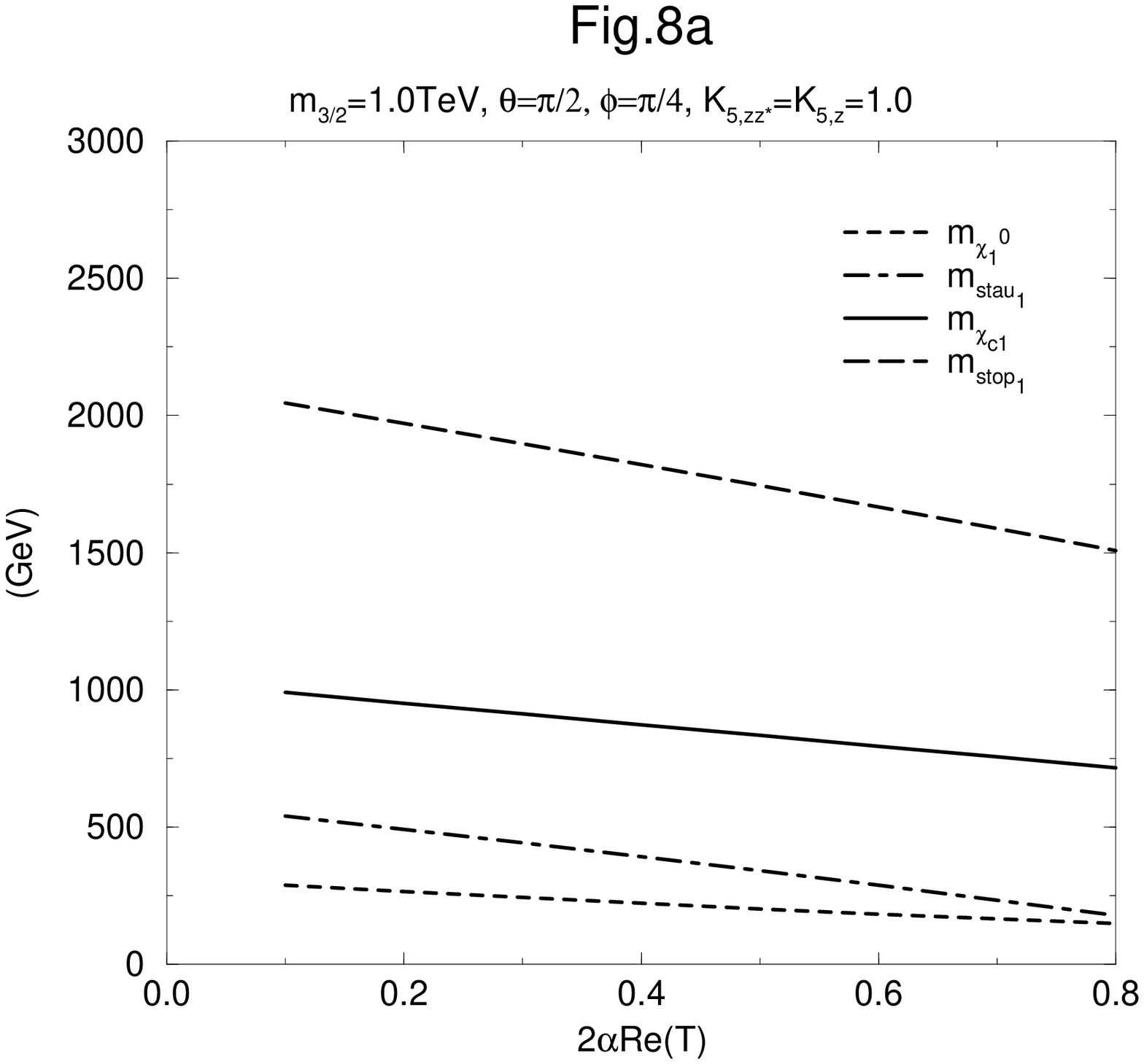}
  \end{center}
 % \caption{}       
 % \label{}       
 \end{minipage}
 \end{figure}

%%%%%%%%%%%%%%%%%%%%%%%%%%%%%%%%%%%%%%%%%%%%%%%%%%%%%%%%%%%%%%%%%%%%%%%%%

 ${}\vspace{-192pt}$

 \begin{figure}[h]     

 ${}\vspace{-80pt}$

 \begin{minipage}{80mm}
  \epsfxsize=110mm     
  \begin{center}
   \leavevmode
   \epsfbox{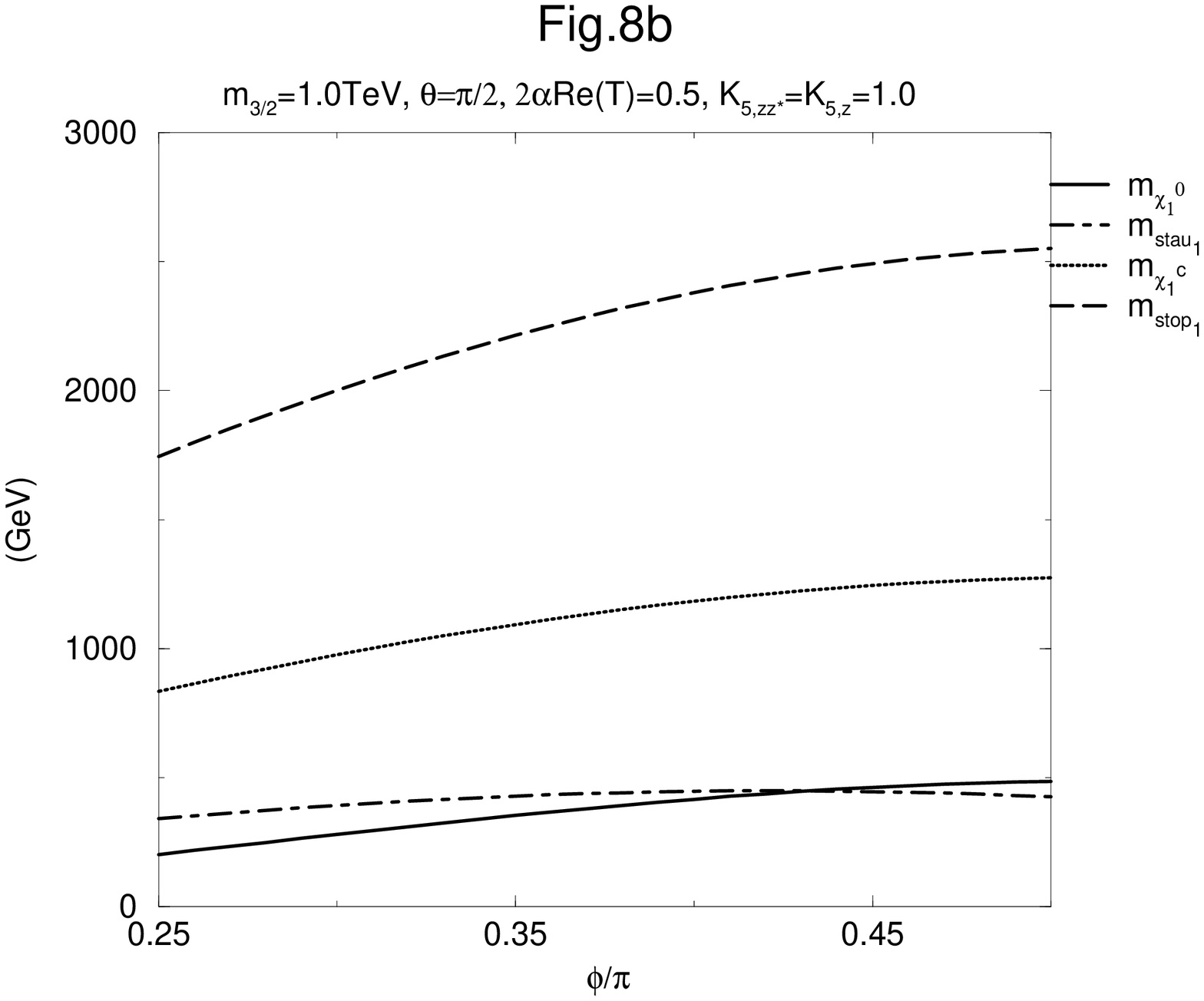}
  \end{center}
 % \caption{}       
 % \label{}       
 \end{minipage}

 ${}\vspace{-122pt}$

 \begin{minipage}{80mm}
  \epsfxsize=110mm  
  \begin{center}
   \leavevmode
   \epsfbox{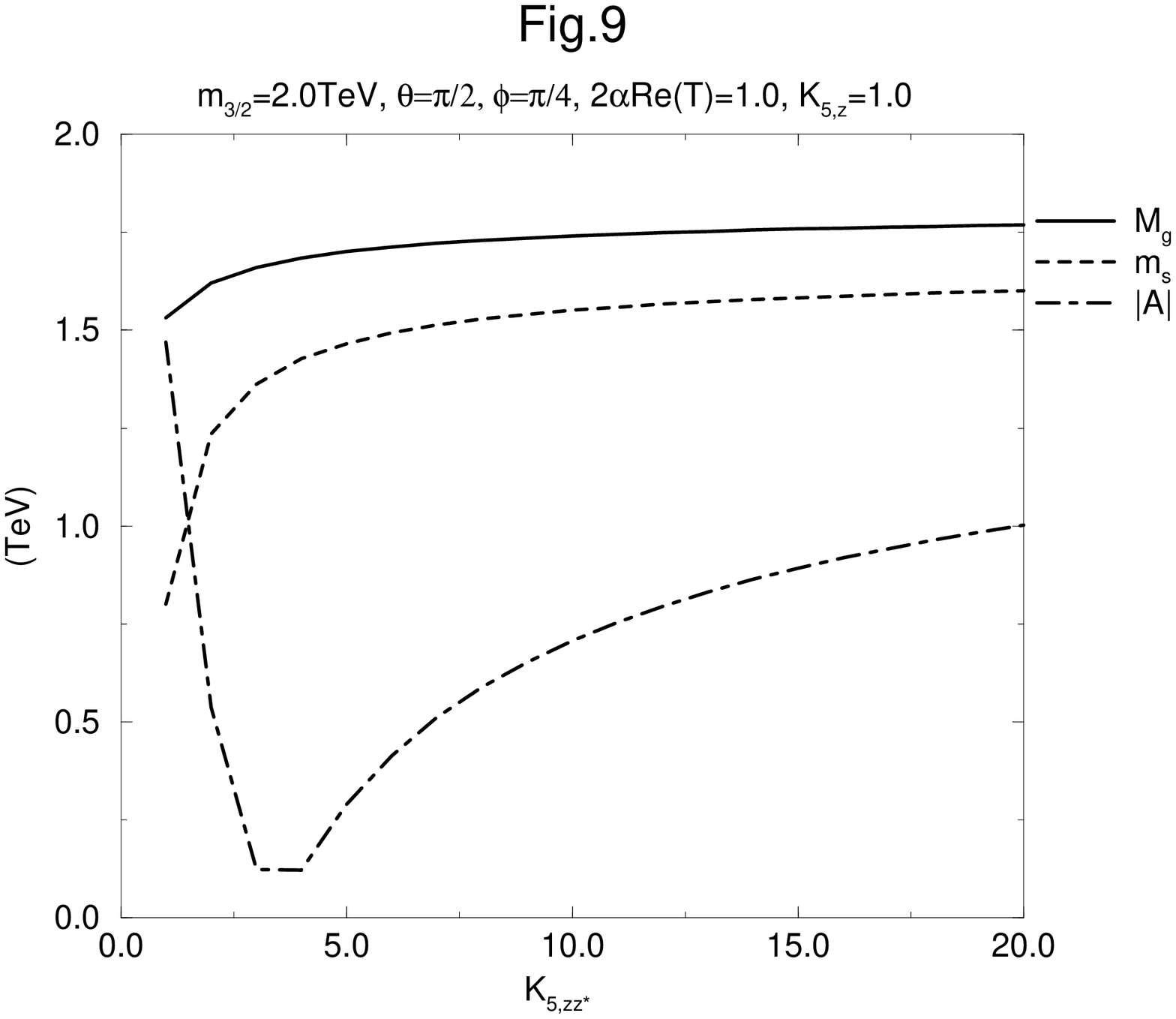}
  \end{center}
 % \caption{}       
 % \label{}       
 \end{minipage}
 \end{figure}

%%%%%%%%%%%%%%%%%%%%%%%%%%%%%%%%%%%%%%%%%%%%%%%%%%%%%%%%%%%%%%%%%%%%%%%%%

 ${}\vspace{-192pt}$

 \begin{figure}[h]     

 ${}\vspace{-80pt}$

 \begin{minipage}{80mm}
  \epsfxsize=110mm     
  \begin{center}
   \leavevmode
   \epsfbox{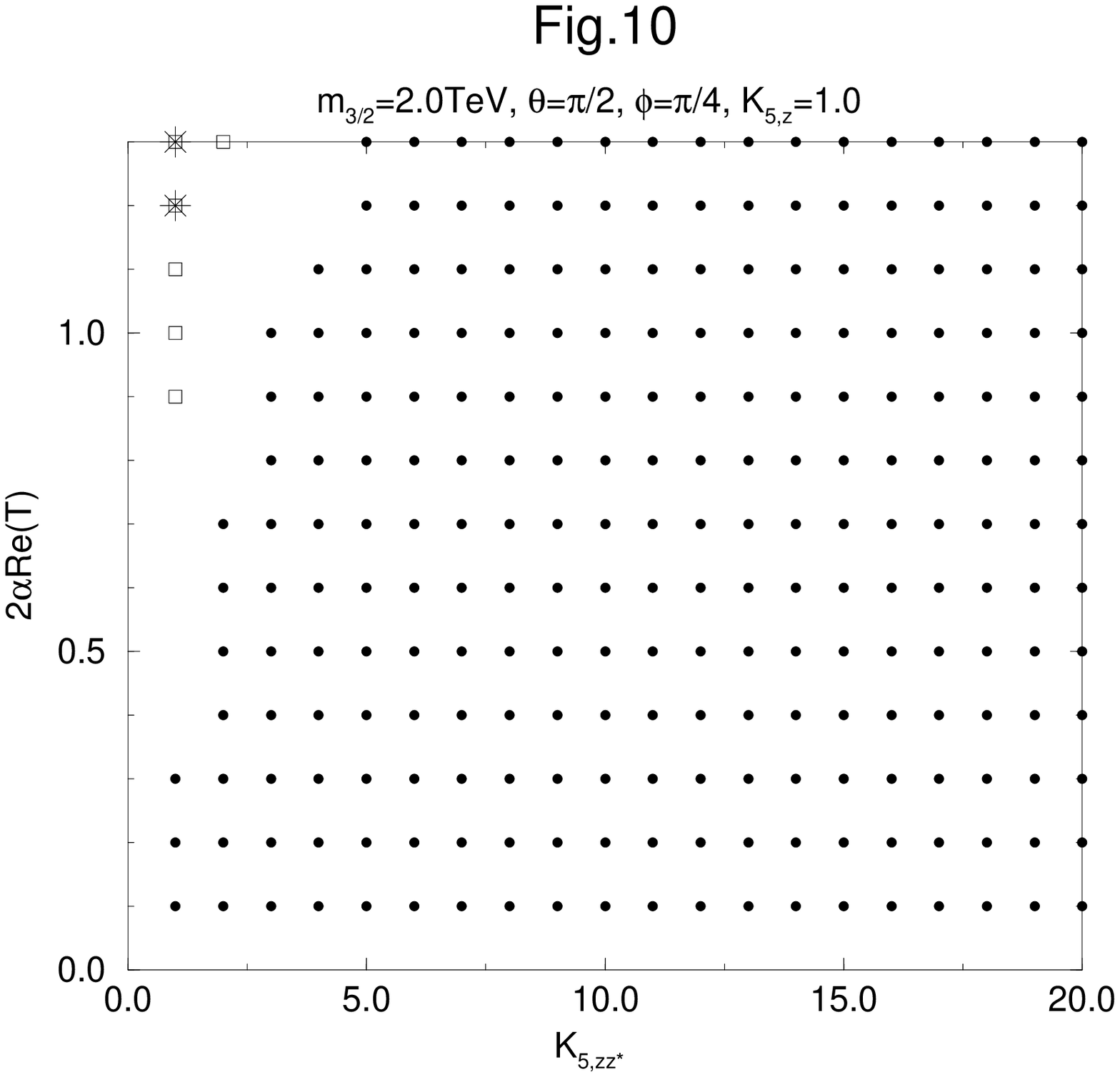}
  \end{center}
 % \caption{}       
 % \label{}       
 \end{minipage}

 ${}\vspace{-122pt}$

 \begin{minipage}{80mm}
  \epsfxsize=110mm  
  \begin{center}
   \leavevmode
   \epsfbox{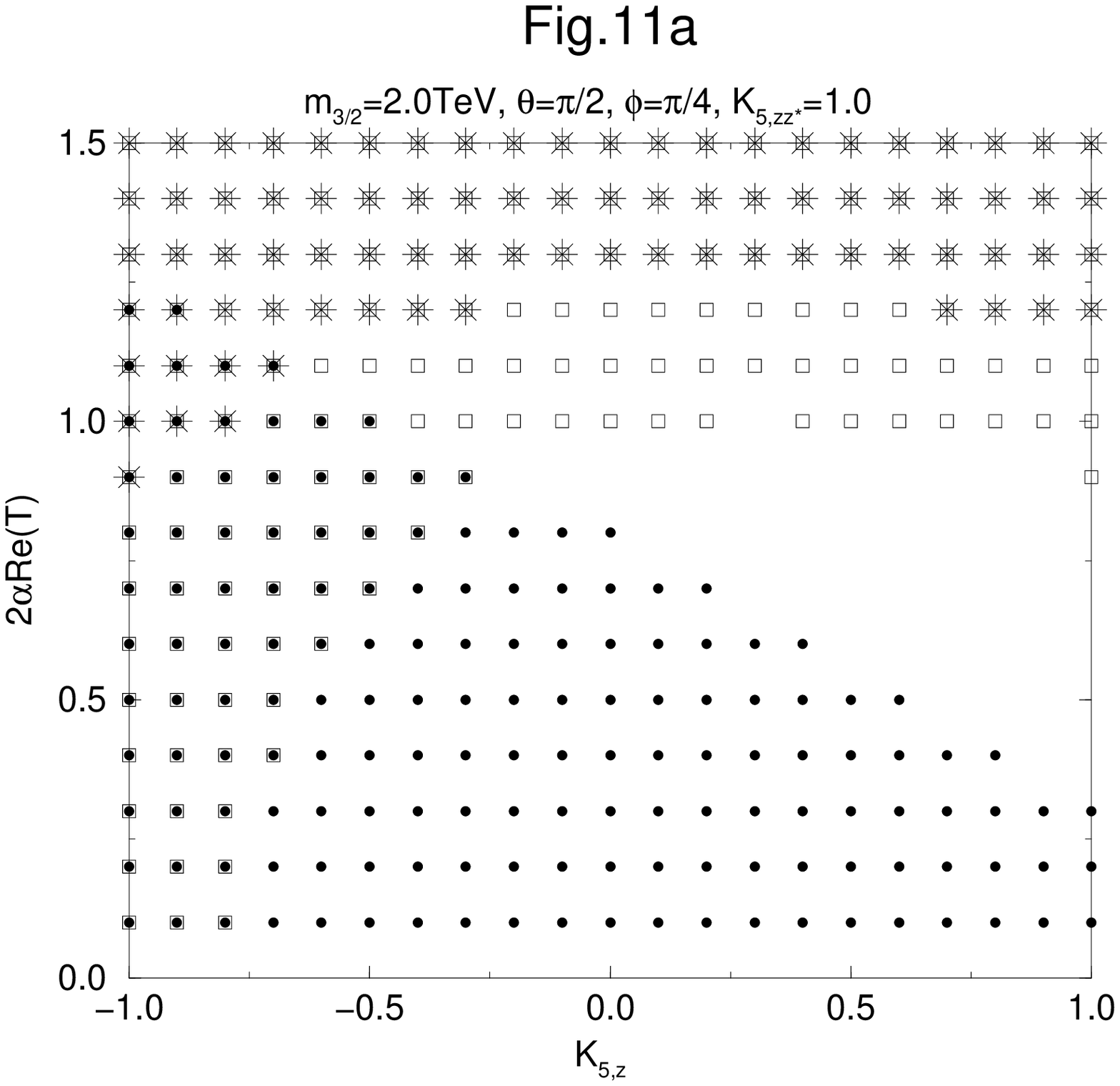}
  \end{center}
 % \caption{}       
 % \label{}       
 \end{minipage}
 \end{figure}

%%%%%%%%%%%%%%%%%%%%%%%%%%%%%%%%%%%%%%%%%%%%%%%%%%%%%%%%%%%%%%%%%%%%%%%%%

 ${}\vspace{-192pt}$

 \begin{figure}[h]     

 ${}\vspace{-80pt}$

 \begin{minipage}{80mm}
  \epsfxsize=110mm     
  \begin{center}
   \leavevmode
   \epsfbox{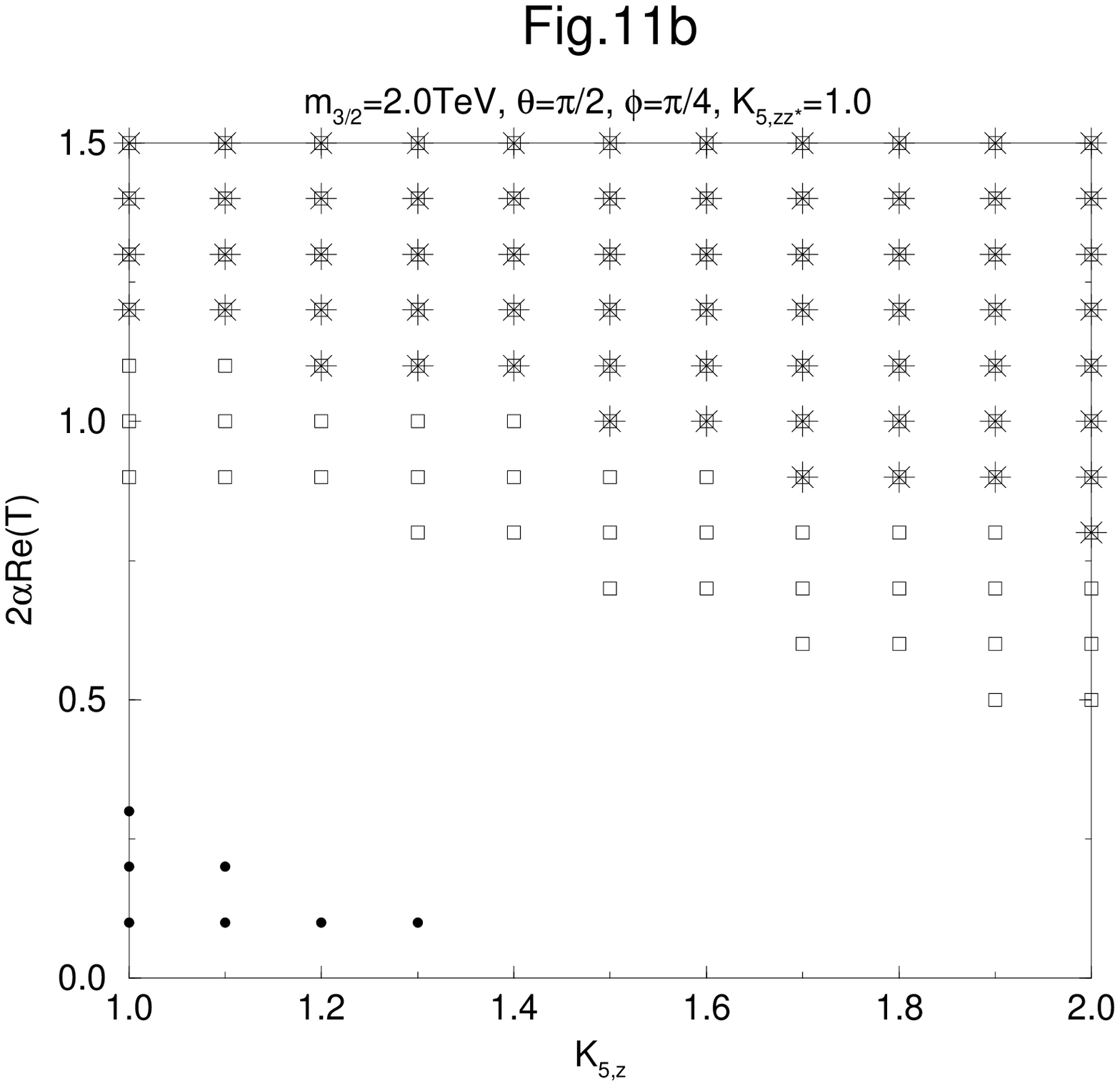}
  \end{center}
 % \caption{}       
 % \label{}       
 \end{minipage}

 ${}\vspace{-122pt}$

 \begin{minipage}{80mm}
  \epsfxsize=110mm  
  \begin{center}
   \leavevmode
   \epsfbox{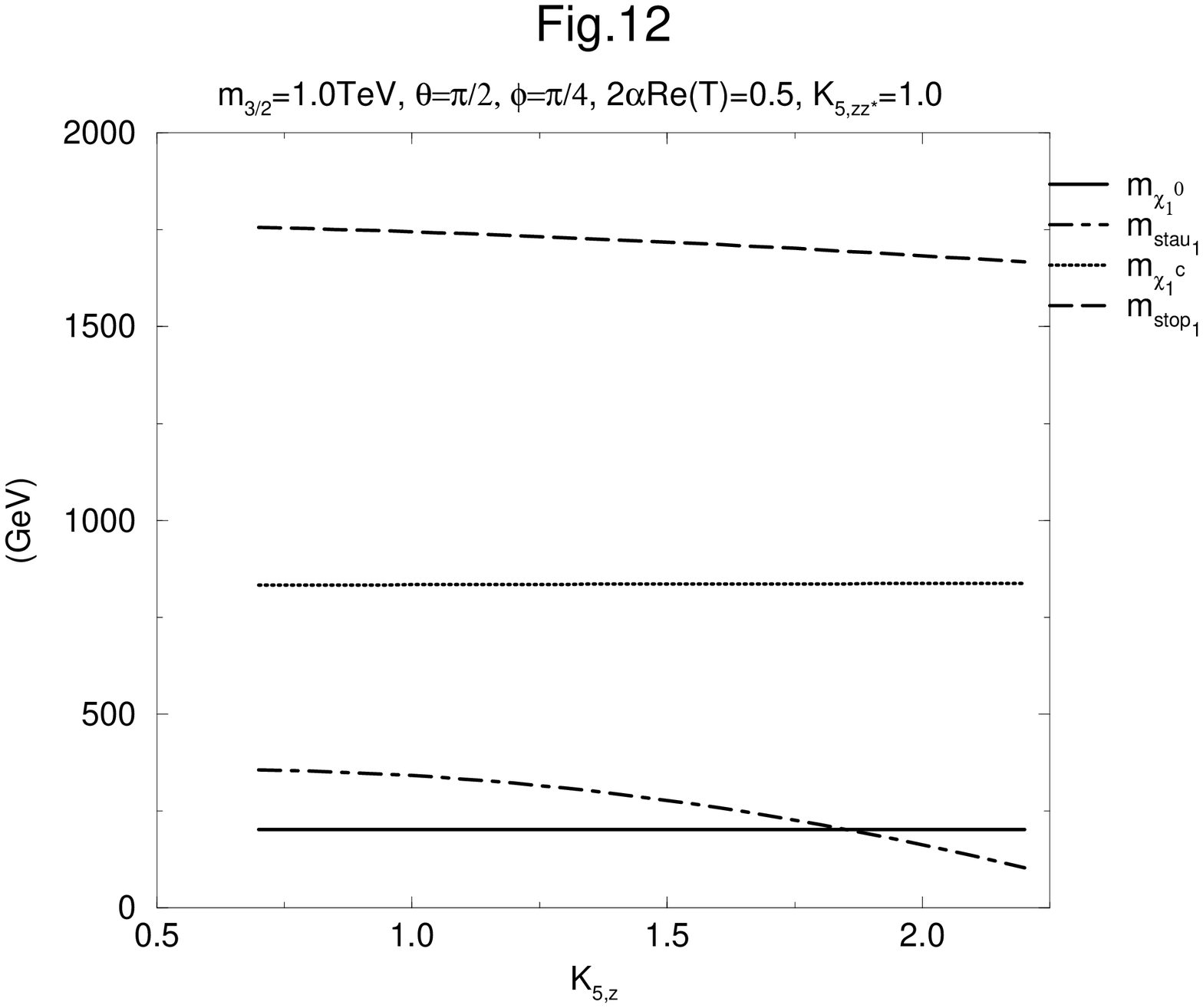}
  \end{center}
 % \caption{}       
 % \label{}       
 \end{minipage}
 \end{figure}

\end{document}